\documentclass[useAMS,usenatbib,fleqn]{mn2e}
\usepackage{times}
\usepackage{amsmath}
\usepackage{graphicx} 
\usepackage{multicol}\usepackage{lastpage}

\usepackage{amssymb}
\usepackage{upgreek}
\usepackage{mathtools}
\usepackage{bm}

\def\d{{\rm d}}
\def\D{{\rm D}}
\def\del{\upartial}

\def\grad_s{\nabla_{\! s}\,}
\def\grad_p{\nabla_{\! p}\,}
\def\bfx{\mathbfit{x}}

\def\bfv{\mathbfit{v}}
\def\bfk{\mathbfit{k}}
\def\frakp{\mathfrak{p}}
\def\bfOmega{\bm{\mathit{\Omega}}}

\def\gsim{\;\rlap{\lower 2.5pt\hbox{$\sim$}}\raise 1.5pt\hbox{$>$}\;}
\def\lsim{\;\rlap{\lower 2.5pt\hbox{$\sim$}}\raise 1.5pt\hbox{$<$}\;}

\def\gsim{\;\rlap{\lower 2.5pt\hbox{$\sim$}}\raise 1.5pt\hbox{$>$}\;}
\def\lsim{\;\rlap{\lower 2.5pt\hbox{$\sim$}}\raise 1.5pt\hbox{$<$}\;}
\def\del{{\upartial}}
\def\grad{\nabla}
\def\bfOmega{\mbox{\boldmath{$\Omega$}}}

\def\bfzeta{\bm{\mathit{\zeta}}}
\def\beq{\begin{equation}}
\def\eeq{\end{equation}}

\newcommand{\LR}{L_{\cal R}}

\usepackage{color}
\definecolor{comred}{rgb}{.8,.2,0.1}
\definecolor{insgreen}{rgb}{.1,.5,0.1}

\title{Modons on Tidally Synchronised Extrasolar Planets}

\author [J.~W.~Skinner \& J.~Y-K.~Cho] {J.~W.~Skinner,$^{1}$\thanks{E-mail:
    j.w.skinner@qmul.ac.uk} J.~Y-K.~Cho,$^{1,2,3}$\thanks{E-mail:
    jcho@flatironinstitute.org} \\ $^1$ School of Physics and Astronomy, Queen
  Mary University of London, Mile End Road, London E1 4NS, UK\\ $^2$ Department
  of Astrophysical Sciences, Princeton University, 4 Ivy Lane, Princeton, NJ,
  08544,USA\thanks{on leave from Queen Mary University of London}\\ $^3$ CCA,
  Flatiron Institute, 162 Fifth Ave, New York, NY, 10010, USA \thanks{present
    address}}

\begin{document}

\date{Accepted yyyy mmm dd. Received yyyy mmm dd; in original form yyyy mmm dd}

\pagerange{\pageref{firstpage}--\pageref{}} \pubyear{yyyy}

\maketitle

\label{firstpage}

\begin{abstract}
We investigate modons on tidally synchronised extrasolar planets.  Modons are
highly dynamic, coherent flow structures composed of a pair of storms with
opposite signs of vorticity.  They are important because they divert flows on
the large-scale; and, powered by the intense irradiation from the host star,
they are planetary-scale sized and exhibit quasi-periodic life-cycles --
chaotically moving around the planet, breaking and reforming many times over
long durations (e.g. thousands of planet days).  Additionally, modons transport
and mix planetary-scale patches of hot and cold air around the planet, leading
to high-amplitude and quasi-periodic signatures in the disc-averaged temperature
flux.  Hence, they induce variations of the “hot spot” longitude to either side of 
the planet’s sub-stellar point – consistent with observations at different epoch.  
The variability behaviour in our simulations broadly
underscores the importance of accurately capturing vortex dynamics in extrasolar
planet atmosphere modelling, particularly in understanding current observations.
\end{abstract}


\section{Introduction}\label{intro}

Modons are translating vortex dipole structures, consisting of two
oppositely-signed patches of vorticity that form in the atmosphere and ocean
\citep{Ster75}.  In atmospheric dynamics, modons play an important role in
establishing ``blocking patterns'' which prevent weather or jet systems from
moving through over a certain region.  In a tidally synchronised planet
atmosphere, a couplet of modons forms -- one composed of a pair of cyclones and
the other composed of a pair of anti-cyclones: cyclones (anti-cyclones) are
storms characterised by vorticity having the same (opposite) sense as the
planetary vorticity \citep[e.g.][]{Holt04}.  These modons are planetary-scale in
size, and they initially span across the equator (assuming zero obliquity for
the planet).  While both modons are spatially compact and isolated, they
generally interact strongly with each other as well as other flow structures
present in the atmospheres of the synchronised planets because of the large
interaction length-scale (the Rossby deformation scale~$\LR$) for the flow
structures.

Currently, observations of hot-Jupiter atmospheres show large variations in the
longitudinal location of the ``hot spot'' as well as the amplitude of spectral
features \citep[e.g.][]{Grill08,Zellem14,Armstrong17,Dang18,Zhang18, Jackson19}; see
 also \citet{Choetal19} for a recent review.  However, atmospheric
flow simulations that use the commonly-employed forcing and initialization
generally produce a large monolithic patch of nearly stationary hot area located
eastward of the sub-stellar point $\sim$2$\times 10^{-3}$\,MPa
\citep{Cooper05,Showmanetal08a}.  The discrepancy between observations and
simulations may arise because simulations thus far have lacked
the required horizontal resolution \citep{SkinCho21}.  Hence the simulations
have not been able to accurately capture the dynamics of small-scales which
influence the large-scale dynamics (e.g. modons and jets) through their
non-linear interactions with the large-scales.  Accurately modelling these
interactions is essential because dynamics serves as the core for other
important atmospheric processes such as radiative transfer, clouds,
photochemistry, and ionization.

In this paper, we describe the results from a set of high-resolution
simulations, which accurately capture the small-scale vortices (storms) and
waves inherent in synchronised planet atmospheres -- hence, accurately represent
large scale dynamics.  Our results are grounded in a series of extensive
numerical convergence and parameter sensitivity studies with the code used in
the present work \citep[][]{PoliCho12,Polietal14,Choetal15,SkinCho21}.  We find
that tidally synchronised planet atmospheres contain a large number of intense
storms, that span a wide range of sizes -- including the planetary-scale.
Significantly, these planetary-scale storms greatly influence the large-scale
spatial distribution and temporal variability of hot, as well as cold, regions
over the planet.  These storms and their motions lead to signatures that may be
observable and, therefore, important for interpreting and guiding current and
future observations \citep{Choetal21}.

The outline of this paper is as follows.  In Section~\ref{method}, we describe
the numerical model we use, how it is set up for the simulations described in
this work and linear theory relevant for the non-linear solutions obtained with
the model.  In Section~\ref{results}, we describe the transient and persistent
solutions that arise when the aforementioned setup, which is commonly-employed
in extrasolar planet studies, is used.  We emphasize that the primary focus of
this paper is on the flow and its associated temperature distributions: results
from a comprehensive study of numerical convergence and accuracy (which includes
the high resolution and dissipation order employed in this work) is reported in
\citet{SkinCho21}, and analysis incorporating radiative transfer, chemistry, and
aerosols will be described elsewhere.  In particular, here we focus on the
quasi-periodic behaviour of modons and their effects on the flow and temperature
fields.  Significantly, this behaviour produces discernible signatures in
disc-averaged temperature fluxes.  In Section~\ref{conclusion}, we conclude by
briefly summarising this work and discussing its implications for extrasolar
planet circulation modelling and observations.

\section{Methodology}\label{method}

\subsection{Governing Equations}

The governing equations, numerical model, and simulation setup
used in this paper are the same as in \citet{SkinCho21} and \citet{Choetal21}.
We reproduce the key points of the equations and model here for the reader's
convenience.  As in many past works, we solve the traditional primitive
equations \citep[see e.g.][]{Salb96} in $\bfx = (\lambda,\phi,p)$ coordinates,
representing (longitude, latitude, pressure) in this work.  The equations read:
\begin{subequations}\label{eq:pe}
  \begin{eqnarray}
    \frac{\D \bfv}{\D t}\ & = & -\grad_p \Phi - \Big(\frac{u}{R_p}\tan\phi+ f
    \Big)\bfk\times\bfv + {\cal D}_{\bfv} \\ \frac{\del\Phi}{\del p}\ & = &
    -\frac{1}{\rho} \\ \frac{\del\omega}{\del p}\ & = & -\grad_p\cdot\bfv
    \\ \frac{\D T}{\D t} & = & \frac{\omega}{\rho\, c_p} +
    \frac{\dot{q}_{\rm{net}}}{c_p} + {\cal D}_T\, ,
  \end{eqnarray}
\end{subequations}
where $\D / \D t \equiv \del / \del t + \bfv\cdot\grad_p + \omega\del / \del p$
is the material derivative; $t$ is the time; $\bfv = (u,v)$ is the (eastward,
northward) velocity (in a frame rotating with rotation rate $\varOmega$) on a
constant $p$-surface; $\omega\equiv\D p / \D t$ is the ``vertical'' pressure
velocity in the rotating frame; $R_p$ is the planetary radius, which is
fiducially set to be at $p\! =\! 0.1$\,MPa; $\bfk$ is the unit vector in the
local vertical direction; $\grad_p$ is the horizontal gradient on a constant
$p$-surface; $\Phi(\bfx,t) = gz(\bfx,t)$ is the geopotential, where $g$ is the
constant surface gravity at $z = R_p$ with $z$ the vertical distance above
$R_p$; $f(\phi) = 2 \varOmega\sin\phi$ is the Coriolis parameter, the projection
of the planetary vorticity vector~$2\bm{\mathit{\Omega}}$ onto~$\bfk$; the
direction of $\bm{\mathit{\Omega}}$ orients north; $T(\bfx,t)$ is the
temperature; $\cal{D}\!_{\chi}$, for $\chi \in \{\bfv, T\}$\footnote{As in
  \citet{SkinCho21}, $\{\,\cdot\,,\,\cdot\,,\,\ldots\}$, $[\,\cdot\,,\cdot\,]$
  and $(\,\cdot\,,\,\cdot\,,\,\ldots)$ carry their usual meanings in this paper
  -- i.e. set, (closed) interval and tuple, respectively.}, is given by
\begin{eqnarray}\label{eq:hyper}
  {\cal D}\!_{\chi}\ =\ \nu_{2 \frakp}\big[(-1)^{\frakp+1}\grad\!_p^{\,2\frakp}
    + {\cal C}\big]\,\chi\, ,
\end{eqnarray}
where $\nu_{2\frakp}$ is the constant dissipation coefficient; $\frakp \in
\mathbb{N}$, where $\mathbb{N} = \{0, 1, 2, \dots\}$, is the order of the
dissipation (not to be confused with the pressure $p$) with $\frakp\! >\! 1$
instantiations known as hyper-dissipation \citep[see
  e.g.][]{ChoPol96a,ThraCho11,PoliCho12}; ${\cal C} = (2/R_p^2)^{\frakp}$ is a
term that compensates the damping of uniform rotation by $\cal{D}\!_{\bfv}$
\citep[see e.g.][]{Polietal14}; $\rho(\bfx,t)$ is the density; $c_p$ is the
constant specific heat at constant pressure; and, $\dot{q}_{\rm{\tiny
    net}}(\bfx,t)$ is the net diabatic heating rate.

Equations~(\ref{eq:pe}) are closed by the equation of state for an ideal gas, $p
= \rho {\cal R}T$, where ${\cal R}$ is the specific gas constant.  A useful
variable is the potential temperature, $\Theta(\bfx,t) \equiv T(P_{\rm
  ref}/p)^\kappa$, where $p_{\rm ref}$ is a constant reference pressure and
$\kappa \equiv {\cal R}/c_p$; for example, $\Theta$ is materially conserved when
$\dot{q}_{\rm net}\! =\! {\cal D}_T = 0$.  The boundary condition for the
equations is ``free-slip'' (i.e. $\D p / \D t = 0$) at the top and bottom
$p$-surfaces; note that the top and bottom boundaries are material surfaces,
across which no mass is transported.  With this boundary condition, the
equations permit the full range of large-scale motions for a stably-stratified,
un-ionized atmosphere -- with the exception of sound waves: sound waves are
filtered out from the full compressible hydrodynamics
equations\footnote{Although sound waves are not admitted, the primitive
  equations are still compressible, as $\D\rho / \D t \ne 0$; see
  equation~(\ref{eq:pe}c).} via the combination of the hydrostatic balance
condition, expressed by equation~(\ref{eq:pe}b), and the free-slip boundary
conditions at the top and bottom.  However, the results presented in this study
(e.g. the emergence of dynamic modons) 
also apply to simulations solving the full Navier-Stokes (non-hydrostatic)
equations employing a similar physical setup, since the speed of fast gravity
waves admitted by both the hydrostatic and non-hydrostatic equations are close
to the speed of the sound wave (see Table~\ref{tab:params}).

\begin{table}
  \caption{Physical, Numerical, and Scale Parameters:\  $^{(a)}$ based on
    $c_p$; $^{(b)}$~for $\{{\rm H}_2, {\rm He}\}$; $^{(c)}$ at $p = 0.1$\,MPa;
    $^{(d)}$ at $p = 1$\,KPa }
  \label{tab:params}
  \begin{tabular}{llll}
    \hline Planetary rotation rate & $\Omega$ & 2.1$\times$10$^{-5}$ 
    & s$^{-1}$ \\ 
    Planetary radius & $R_p$ & 10$^8$ & m \\ 
    Surface gravity & $g$ & 10 & m\,s$^{-2}$ \\ 
    Specific heat at constant $p$ & $c_p$ & 1.23$\times$10$^4$ 
    & J\,kg$^{-1}$\,K$^{-1}$ \\ 
    Specific gas constant$^{(a,b)}$  & ${\cal R}$ & 3.5$\times$10$^3$ 
    & J\,kg$^{-1}$\,K$^{-1}$ \\ 
    \\ 
    Initial temperature$^{(c)}$ & $T_m$ & 1600 & K \\ 
    ``Equil.'' sub-stellar temp.$^{(c)}$ & $T_{e_{\rm d}}$ & 1720 & K \\
    ``Equil.'' anti-stellar temp.$^{(c)}$ & $T_{e_{\rm n}}$ & 1480 & K \\
    Thermal relax. time $^{(d)}$ & $\tau_{\rm th}$ &  $\approx\! 10^5$ & s \\  
    Pressure at top & $p_{\rm top}$ & 0 & MPa \\
    Pressure at bottom & $p_{\rm bot}$ & $[0.1,10]$ & MPa \\     
    Pressure w/o forcing   & $p_0$ & $ \ge 1 $ & MPa \\
    \\
    Truncation wavenumber & T & $[21,682]$ & \\ 
    Number of levels (or layers) & L & $[3,1000]$ & 
    \\ Max. sectoral wavenumber & $M$ & $= {\rm T}$ &
    \\ Max. total wavenumber & $N$ & $= {\rm T}$ & \\ 
    Dissipation operator order & $\frakp$ & $[1,8]$ & \\
    Viscosity coefficient & $\nu_{2\frakp}$ & (see text) &
    m$^{2\frakp}$\,s$^{-1}$\\
    (Hyper)dissip. wavenumber & $n_{d(2\frakp)}$ & (see text) & \\ 
    \\
    Vertical length scale & ${\cal H}$ & $\sim\! {\cal R}T_m / g$ & m \\
    Horizontal length scale & ${\cal L}$ & $\!\ga R_p / 20$ & m \\
    Maximum jet speed & ${\cal U}$ & $\!\ga 2\!\times\! 10^3$ & m\,s$^{-1}$ \\
    Sound speed$^{(c)}$ & $c_s$ & $\approx\! 2.8\!\times\! 10^3$ & m\,s$^{-1}$ \\ 
    Dissipation time-scale & $\tau_d$ & $\sim\! 2\!\times\! 10^5$ & s \\ 
    Brunt-V\"ais\"al\"a frequency & ${\cal N}$ & $\sim\! 2\!\times\! 10^{-3}$ 
    & s$^{-1}$ \\ 
    Rossby number & $R_{\rm o}$ & $\equiv {\cal U} / (\Omega{\cal L})$ &  \\ 
    Froude number & $F_{\rm r}$ & $\equiv {\cal U} / \sqrt{g{\cal H}}$ & \\ 
    Rossby deformation scale & ${\cal L}_{\cal R}$ & $\equiv\sqrt{g{\cal H}}/\Omega$ & m \\
    \hline\\
    \end{tabular}
\end{table}

\subsection{Numerical Model} 

We solve equations~(\ref{eq:pe}) and (\ref{eq:hyper}) numerically using the
pseudospectral code, BOB \citep[][]{Rivietal02,Scotetal04}.  BOB is a
highly-accurate ``dynamical core'' of a general circulation model (GCM).  GCMs
are typically used in atmospheric dynamics studies and climate modelling of the
Solar System planets.  But, BOB has been rigorously tested and validated under
the numerically stringent conditions typical of hot-Jupiters; see, for example,
\citet{PoliCho12}, \citet{Polietal14}, \citet{Choetal15} and \citet{SkinCho21}.

BOB solves the equations in the ``vorticity-divergence and potential
temperature'' form\footnote{the curl and divergence of equation~(\ref{eq:pe}a),
  along with equation~(\ref{eq:pe}d) in terms of the potential temperature}.  In
this form, equations~(\ref{eq:pe}) are more suited to the spectral transform
method \citep[see e.g.][]{Orsz70,Eliaetal70,Canuto88}, which offers superior
convergence properties compared to the traditional (e.g. finite difference)
schemes \citep[e.g.][]{Boyd00,Durr10}.  BOB is essentially a multi-layer
extension of the 1-layer codes used in the studies of Solar System giant planets
by \citet{ChoPol96a,ChoPol96b} and extrasolar system giant planets by
\citet{Choetal03,Choetal08}.  The time integration of the equations in all of
these codes is performed using a second-order accurate, leap-frog scheme with a
small amount of Robert--Asselin filter applied to suppress the computational
mode arising from the scheme \citep{Robe66,Asse72}.  The time-step size $\Delta
t$ in all the simulations are such that the Courant-Friedrichs-Lewy (CFL) number
\citep[e.g.][]{Stri04,Durr10} is well below unity -- typically $ <\! 0.3$.

For each $p$-surface, the code transforms the equations to the spectral space
with a ``triangular truncation'' -- i.e. up to $N\!  =\! M \equiv {\rm T}$
wavenumbers retained in the Legendre expansion,
\begin{equation}
  \xi(\lambda,\mu,t)\ =\ \sum^N_{n = 0}\ \sum^M_{\ m = -M} \xi^m_n(t)\,
  Y^m_n(\mu,\lambda), \quad |m| \le n\,,
\end{equation}
where $\xi$ is an arbitrary scalar field; $\mu\!\equiv\!\sin\phi$; $n \in
\mathbb{N}$ and $m \in \mathbb{Z}$ are the total and sectoral wavenumbers,
respectively; $(N,M) \in \mathbb{N}^2$; $Y^m_n(\lambda,\mu) \equiv
P^m_n(\mu)\,e^{i m \lambda}$ are the spherical harmonic functions; and, $P^m_n$
are the associated Legendre functions.  The set $\{Y^m_n\}$ are the
eigenfunctions of the spherical Laplacian operator:
\begin{equation}
  \grad^2\,Y^m_n\ =\ -\!\left[\frac{n(n + 1)}{R_p^2}\right]Y^m_n\, ,
\end{equation}
where
\begin{equation}
  \grad^2\ =\ \frac{1}{R_p^2}\left\{\frac{\del}{\del\mu}
    \left[\left(1 - \mu^2 \right)\frac{\del}{\del\mu}\right]\, +\, 
    \frac{1}{1 - \mu^2}\frac{\del^2}{\del\lambda^2}\right\}\, .
\end{equation}
The $\{Y^m_n\}$ constitutes a complete, orthogonal expansion basis
\citep[e.g.][]{ByroFull92}.  Note that, when $\frakp = 1$,
equation~(\ref{eq:hyper}) reduces to the Laplacian operator acting on $\chi$,
modulo ${\cal C}$.  Note also that a representation in spectral space with a
truncation wavenumber~T is transformed to a Gaussian grid in physical space with
approximately $(3{\rm T},3{\rm T}/2)$ points in the $(\lambda,\phi)$-direction.
However, the Gaussian grid should {\it not} be directly compared with the grid
of a finite-difference (or other grid-based) methods, as the former grid is
effectively equivalent to a much higher resolution than a finite-difference grid
with the same number of points.  This is due to the pseudospectral method's
accuracy and convergence properties and the use of $\frakp \gg 1$
\cite[e.g.][]{ChoPol96a,Boyd00,Durr10,SkinCho21}.

Since our goal is to follow the evolution of highly dynamic and minimally
dissipated flow structures over a long duration, we use $\mathfrak{p} = 8$.  The
role of $\cal{D}_{\chi}$ is to limit dissipation to the small-scales and to
provide a conduit for energy and enstrophy cascade that prevent the simulation
from ``blowing up''.  As has been demonstrated in \cite{SkinCho21},
$\mathfrak{p} \geq 8$ is required at the currently practical resolutions to
adequately capture the full range of behaviours exhibited in the flow.  Using a
lower $\mathfrak{p}$, especially at low resolution, can dissipate large-scale
structures -- even the $n\! =\! 2$ mode structures, important in the present
work.  Other than the very weak Robert–Asselin filter to separate out the
computational mode arising from the leapfrog scheme \citep{ThraCho11}, no other
numerical dissipators, drags, fixers, stabilisers or filters are used in
performing the simulations, as they are not necessary in our code.

Vertically, the domain is decomposed into ${\rm L}\in\mathbb{Z}^+$
uniformly-spaced points or layers in the $p$-coordinate.  Along this direction,
a second-order finite-difference scheme is used -- as is common in codes solving
equations~(\ref{eq:pe}) \citep[e.g.][]{Durr10}.  Given the range, $p \in [p_{\rm
    top},p_{\rm bot}]$, the dynamically active levels $p_k$ for $k \in [1,{\rm
    L}]$ are located at
\begin{equation}
  p_k\ =\ \Big(k - \frac{1}{2}\Big)\Big[\frac{p_{\rm bot} - p_{\rm top}}{{\rm
        L}}\Big]\, .
\end{equation}
As already mentioned, the bounding surfaces, $p_{\rm top}$ and $p_{\rm bot}$,
are not dynamically active; but, they enforce the boundary conditions.  Note
that many studies employ a $\log(p)$-spacing \citep[e.g.][]{LiuShow13}; but,
they tend to solve the equations in a {\it uniform} $p$-coordinate, which
introduces a numerical complication \citep{Choetal15}.  This difference in the
vertical spacing, however, does not alter the main results of the present paper.

\subsection{Simulation Setup}\label{setup}

The physical parameters and their values for equation~(\ref{eq:pe}) are shown in
Table~(\ref{tab:params}).  The parameter values are representative of the
tidally synchronised extrasolar giant planet {\it HD209458b}.  Note that the
values in the table are identical to those used in many hot-Jupiter modelling
studies \citep[see e.g.][]{ThraCho10, SkinCho21}.  This facilitates equatable
comparisons.  Note also that, due to the free-slip boundary condition at $p_{\rm
  bot}$, the results in this paper are relevant to telluric planets with a solid
or liquid surface, away from their boundary layers. \footnote{
    Here, by `telluric' we mean a planet with a solid lower boundary (and
    smaller radius).}

\begin{figure} %
  \centerline{\includegraphics[scale=.09]{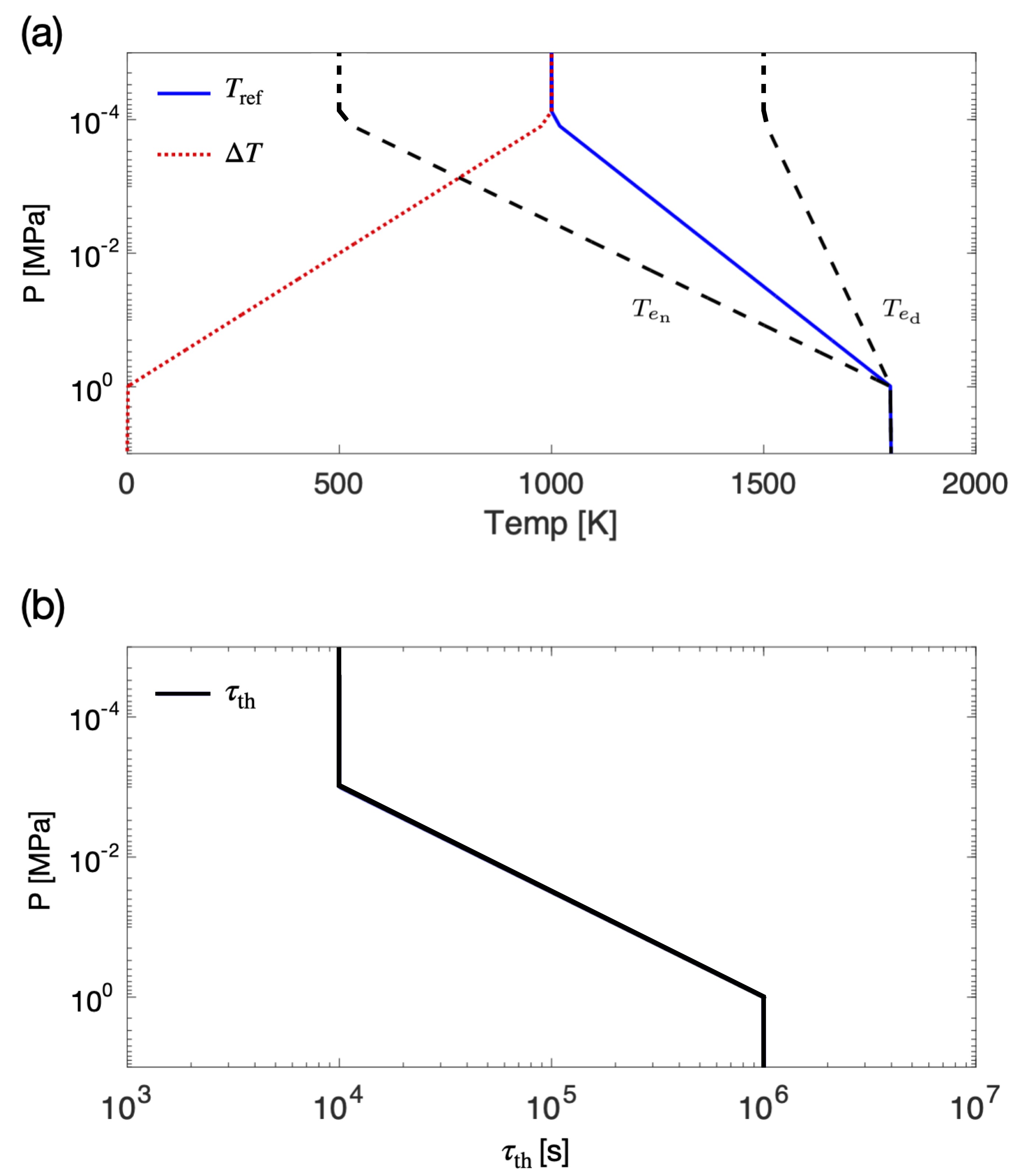}}
  \caption{ Temperature~(a) and relaxation time~(b) profiles used in the
    Newtonian cooling scheme.  (a)~Blue solid line shows the reference
    temperature $T_{\rm ref}$, which is the initial temperature (uniform at each
    $p$).  The accompanying black dashed lines show the equilibrium temperature
    profiles, $T_{e_{\rm d}}$ at the sub-stellar point (right) and $T_{e_{\rm
        n}}$ at the anti-stellar point (left); the latter temperature is a
    constant over the entire night-side at each $p$-level.  The red dotted line
    is the equilibrium temperature difference, $\Delta T \equiv T_{e_{\rm d}} -
    T_{e_{\rm n}}$.  (b)~The thermal relaxation time $\tau_{\rm th}$ is roughly
    proportional to $p/T_{\rm ref}^4$ in the sloping region ($p =
    [10^{-3},1]$\,MPa) and uniform above and below ($\tau_{\rm th}$ is a
    constant at each $p$-level).  Outside the sloping region, $\tau_{\rm th}$ is
    very short at the top ($\tau_{\rm th} \ll \tau_{\rm ad}$ at $p \le
    10^{-3}$\,MPa) and long at the bottom ($\tau_{\rm th} \gg \tau_{\rm ad}$ at
    $p \ge 1$\,MPa). }
  \label{fig:prof1}
\end{figure}

For the thermal forcing of the planet's atmosphere, we adopt a commonly utilised
scheme that has been implemented in many past studies \citep[see
  e.g. ][]{Showman02,Choetal03,Cooper05,Choetal08,Showmanetal08a,Menou09,
  Rauscher10,ThraCho10,Hengetal11,LiuShow13,Choetal15,SkinCho21}.  In this
idealised setup, an atmosphere which is initially at rest is driven by a thermal
relaxation to a prescribed ``equilibrium temperature''.  Specifically,
$\dot{q}_{\rm{\tiny net}}/{c_p}$ in equation~(\ref{eq:pe}d) is set to be $-(T -
T_{\rm eq}) / \tau_{\rm th}$, where $T_{\rm eq} = T_{\rm eq}(\lambda, \phi, p)$
is the equilibrium temperature and $\tau_{\rm th} = \tau_{\rm th}(p)$ is the
radiative ``cooling'' time \footnote{The ``spin up'' time varies
    with depth and vertical resolution in all simulations, but it is around 10
    planetary days for the $p_{\rm bot} \sim 0.1$ simulations. For $p_{\rm bot}
    \ge 1$, ``spin up'' time is undefined \citep{Mend20}.}; more precisely, it
is the thermal relaxation time.  The $T_{\rm eq}$ distribution is specified as:
\begin{eqnarray}
  T_\text{eq}\ \ =\ \
\begin{cases}
  T_{e_{\rm n}}(p) + \Delta T(p)\cos\lambda\cos\phi, & \quad
  \text{day-side} \\
  T_{e_{\rm n}}(p), & \quad \text{night-side}\, .
\end{cases}
\label{eq:teq}
\end{eqnarray}
Here $T_{e_{\rm n}}$ is the equilibrium temperature at the night-side, which is
uniform at each $p$.  In contrast, the day-side equilibrium temperature is not
uniform, with $T_{e_{\rm n}} + \Delta T \equiv T_{e_{\rm d}}(p)$ the temperature
at the sub-stellar point, $(\lambda, \phi) = (0^{\circ},0^{\circ})$, for $\Delta
T \equiv T_{e_{\rm d}} - T_{e_{\rm n}}$.  The specified, simple profiles
$\{T_{\rm ref}, \Delta T, T_{e_{\rm n}}, T_{e_{\rm d}}, \tau_{\rm th}\}$ all
depend only on $p$ and crudely represent the effects of irradiation from the
planet's host star on the dynamics.  The profiles are shown in
Fig.~\ref{fig:prof1}.

In Fig.~\ref{fig:prof1}, each of the profiles are piece-wise continuous
functions of $p$ that span the vertical range, $p =[10^{-5}, 10^1]$\,MPa.  It is
important to note that for this setup the strength of the thermal forcing is
{\it flow independent}, which is not realistic (especially when $\tau_{\rm th}
\la \tau_{\rm ad}$, where $\tau_{\rm ad} \equiv {\cal L} / {\cal U}$ with ${\cal
  L} = \pi R_p$ is the advective time).  Moreover, the unknown $\tau_{\rm
  th}(p)$ is chosen so that the forcing strength is monotonic and strongest at
the top of the atmosphere and goes to zero at $p = 1$~MPa; at $p \ge 1$\,MPa,
forcing is not applied.  Note that quantitative aspects of the flow are affected
by the chosen profiles (as well as by the initial flow state) -- as emphasized
repeatedly in the past \citep[][]{Choetal08,ThraCho10,Choetal15,Choetal19}.
Another feature to note is the zonal\footnote{``Zonal'' refers to eastward and
  ``meridional'' refers to northward (and sometimes more loosely east--west and
  north--south, respectively).} {\it a}symmetry of $T_{\rm eq}$, which is
responsible for generating the main (zonally-asymmetric) flow structure
discussed in this paper: the cyclonic and anti-cyclonic modon pair.  ``Deeper
asymmetry'' about the terminator in $T_{\rm eq}$, such as the $\cos\lambda$
distribution extending to the nightside \citep[e.g.][]{ThraCho10,ShowPol11},
leads to stronger asymmetry in the flow.  The effects of the asymmetry are
discussed more in detail in section~\ref{results}.

\begin{figure*} %
  \centerline{\includegraphics[scale=.34]{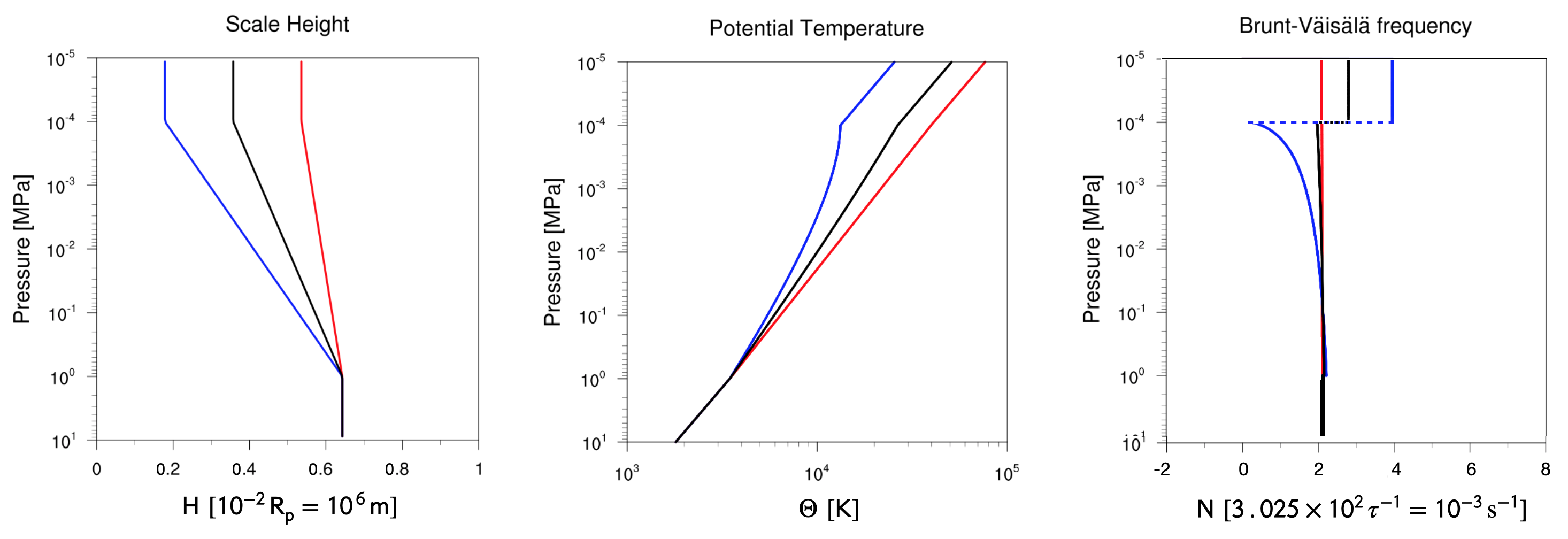}}
  \caption{ Level mean scale height $H(p)$ in units of Mm~(left), potential
    temperature $\Theta(p)$ in units of K~(centre), and
    Brunt-V\"{a}is\"{a}l\"{a} frequency $N(p)$ in units
    of~$10^{-3}$\,s$^{-1}$~(right) for the temperature profiles in
    Fig.~\ref{fig:prof1}.  The black lines correspond to the initial profile
    $T_{\rm ref}$; and, the blue and red lines correspond to equilibrium
    profiles of the anti-stellar and sub-stellar points ($T_{e_{\rm n}}$ and
    $T_{e_{\rm d}}$), respectively.  The temperature profiles lead to abrupt
    changes in the vertical $\Theta$ gradients -- i.e. jumps in the basic
    stratification -- as seen in the Brunt-V\"{a}is\"{a}l\"{a} frequencies.  In
    the latter, the blue profile nearly vanishes at $p \approx 10^{-4}$\,MPa,
    but it is still positive. }
  \label{fig:prof2}
\end{figure*}

Fig.~\ref{fig:prof2} shows three useful profiles derived from the temperature
profiles presented in Fig.~\ref{fig:prof1}: level mean atmospheric scale height,
$H(p) \equiv ({\cal R}\int_k^{k+1} T\, \d\ln p)\, /\,(g\int_k^{k+1} \d\ln p)$,
in units of Mm (left); potential temperature, $\Theta(p) \equiv T(10/p)^\kappa$,
in units of~K (centre); and, Brunt-V\"{a}is\"{a}l\"{a} frequency, $N(p) \equiv
[-\rho g^2(\d\ln\Theta/\d p)]^{1/2}$, in units of~$10^{-3}$\,s$^{-1}$ (right).
The black lines correspond to the initial profile $T_{\rm ref}$; and, the blue
and red lines correspond to equilibrium profiles $T_{\rm eq}$ of the
anti-stellar and sub-stellar points, respectively.  The temperature profiles lead
to abrupt changes in the vertical $\Theta$ gradients -- i.e. jumps in the basic
stratification -- as seen in the Brunt-V\"{a}is\"{a}l\"{a} frequencies.  Note,
the blue profile nearly vanishes at $p \approx 10^{-4}$\,MPa, but it is still
positive.  It is also an artefact of the unrealistic discontinuity in the
corresponding temperature profile.  Formally, the hydrostatic balance condition
restricts the validly of equations~(\ref{eq:pe}) to the stably-stratified
radiative region $p \lsim 10$\,MPa, which overlies an unstably-stratified
convective region $p \gsim 10^2$\,MPa.  The exact transition level between the
stable and unstable region is uncertain, and likely laterally dependent.

From hereon, {\it the planetary radius $R_p$ and rotation period $\tau$
  ($\equiv~2\pi/\Omega = 3.025\!\times\!  10^5$\,{\rm s}) are used as the length
  and time scales}, respectively -- whenever confusion is not
engendered. That is, $\nu_{2\frakp}$ is given in the units of
$R_p^{2\frakp}\,\tau^{-1}$ and $n$ is given in the units of $R_p^{-1}$, by
default.  However, {\it the temperature $T$ and pressure $p$ are given the units
  of {\rm K} and {\rm MPa}}, respectively for easier comparison
  with past simulation and observatonal studies.  Hence, for all the
simulations discussed in this paper, $p_{\rm top} = 0$ while $p_{\rm bot} \in
\{0.1, 1.0, 10\}$.  Note that a range of simulation domains are chosen to
explore the robust features, for the given setup.  Appropriate $p$-levels for
the top and bottom of the simulations are currently unknown \citep{Choetal08}.

\subsection{Linear Theory}\label{linear}

Before embarking on the full non-linear solutions, a brief discussion of linear
theory is instructive.  Under the barotropic (vertically aligned) and isochoric
(constant density) assumptions, equation~(\ref{eq:pe}) reduces to the
shallow-water equations for a single layer of fluid \citep{Pedl87}.  Here we
confine our attention to a tangent plane situated at the equator ($\phi = 0$),
known as the equatorial $\beta$-plane approximation
\citep[e.g.][]{Mats66,Gill80,Wuetal01}:
\begin{subequations}\label{eq:swe} 
\begin{eqnarray}
    \frac{\D \bfv}{\D t}\ & = & -g\grad h - \beta y \bfk \times \bfv\, ,
    \\ \frac{\D h}{\D t} & = & -h\grad\!\cdot\! \bfv\, ,
\end{eqnarray}
\end{subequations}
where now $\D / \D t \equiv \del / \del t + \bfv\!\cdot\!\grad$ with $\grad
\equiv (\del / \del x, \del / \del y)$; $\bfv = (u,v) \in \mathbb{R}^2$, where
$u(\bfx,t)$ is the zonal velocity and $v(\bfx,t)$ is the meridional velocity;
$h(\bfx,t)$ is the fluid thickness; $\beta \equiv \d f / \d y\, \vert_{y = 0} =
2\varOmega / R_p$ is the meridional gradient of the Coriolis parameter at the
equator; and, $\bfk$ is again the unit vector in the vertical direction.  With
the boundary condition, $y \rightarrow \pm \infty$, the dynamics is effectively
confined in the equatorial region because of the finiteness of $\LR$; hence, the
equatorial beta plane acts like a wave guide along the equator, with a number of
different zonally propagating waves that are confined in the meridional
direction and divided into fast and slow types.  Importantly, equatorial waves
in equations~(\ref{eq:pe}) have the same horizontal structure as those admitted
by equations~(\ref{eq:swe}).  However, we stress that the use of $\beta$-plane
model here is used to motivate the explanation for the initial formation of the
modon (only).  The Matsuno-Gill solution for the equatorial $\beta$-plane
\citep[e.g.][]{ShowPol11}, for example, does {\it not} formally and
realistically apply to hot synchronised planet atmospheres\footnote{particularly
  to explain the supersonic, zonally-symmetric, equatorial jet which develop
  after a long time in low-resolution simulations} because of the following
\citep{Choetal19}: {\it i)}~large meridional scale of the modons invalidates the
tangent plane approximation; {\it ii)}~a pre-existing, strongly
  time-varying background flow and temperature field due to dynamic modons and
other storms, which preclude classical eddy--mean flow interaction theory
\citep[e.g.][]{Vall17}; {\it iii)} incompressible (small Mach number) and
homogeneous-layer (unconstrained $\LR$) assumptions of the shallow-water model;
 and, {\it iv)} hot-Jupiters are not expected to have a linear
  Rayleigh drag.

Upon non-dimensionalising equations~(\ref{eq:swe}) with the scaling, $(\bfv,\bfx,t)
\rightarrow (U\bfv,L\bfx,\tau_s t)$, where $U$, $L$, and $\tau_s \equiv L/U$ are
the characteristic speed, length, and time scales, respectively, and then
linearising the equations to zeroth-order in the Rossby number, $R_o \equiv U /
(\beta L^2)$, we obtain:
\begin{subequations}\label{eq:nswe} 
\begin{eqnarray}
    \frac{\del u}{\del t} - vy + \frac{\del h}{\del x}\ & = & 0\, , \\ \frac{\del
      v}{\del t} +uy + \frac{\del h}{\del y}\ & = & 0\, , \\ \frac{\del h}{\del
      t} + \frac{\del u}{\del x} + \frac{\del v}{\del y} & = & 0\, .
\end{eqnarray}
\end{subequations}
From this, we can find a general solution for linear modes at the equator, by
eliminating $u$ and $h$ and obtaining an equation for $v$:
\begin{equation}
  \frac{\del\ }{\del t}\Big(\nabla^2 v - y^2 v - \frac{\del^2 v}{\del t^2}\Big)
  + \frac{\del v}{\del x}\ =\ 0\, .
\end{equation}
Now, expanding $v$ in a series in the Hermite polynomial basis $\varphi_l$
\citep{AbraSteg65}, $v = \sum_l\, v_l(x,t)\varphi_l(y)$, and using 
\begin{equation}
  \frac{\d^2\varphi_l}{\d y^2} + (2l -y^2 +1)\,\varphi_l\ =\ 0
\end{equation}
gives the following equation:
\begin{equation}
  \frac{\del\ }{\del t}\Big[\frac{\del^2 v_l}{\del x^2} - (2l+1) v_l -
    \frac{\del^2 v_l}{\del t^2}\Big] + \frac{\del v_l}{\del x} = 0\, .
\end{equation}
This leads to
\begin{equation}\label{eq:3rdo}
  \frac{\del^3 \hat{v}_l}{\del t^3} + (k^2 + 2l + 1)\frac{\del\hat{v}_l}{\del t}
  - ik\hat{v}_l = 0\, ,
\end{equation}
where $\hat{v}_l(k,t)\ =\ \int v_l(x,t)\, e^{ikx}\, \d x + c.c.$ is the Fourier
transform of $v_l$.  The general solution for equation~(\ref{eq:3rdo}) is:
\begin{equation}
  \hat{v}_l = \sum_{\alpha=1}^3\, v_\alpha(k)\, e^{-i\omega_\alpha t}\, ,
\end{equation}  
where $\omega_\alpha$ for $\alpha = 1, 2, 3$ are the three roots of the
dispersion equation:
\begin{equation}\label{eq:disper}
  \omega_\alpha^3 - (k^2 + 2l +1)\omega_\alpha - k\ =\ 0\, .
\end{equation}  
Here, the lowest value of $\omega_\alpha$ at a given $(k,l)$ corresponds to a
wave with westward propagation.  This is the equatorial Rossby wave, important
for the large-scale structure in our work.  Note that $l$ gives the number of
nodes in the meridional direction for $v$.  The approximate solution to
equation~(\ref{eq:disper}) is
\begin{equation}
  \omega_l(k)\ =\ \frac{k}{2l + k^2 + 1}\, .
\end{equation}

\begin{figure} %
  \centerline{\includegraphics[scale=.23]{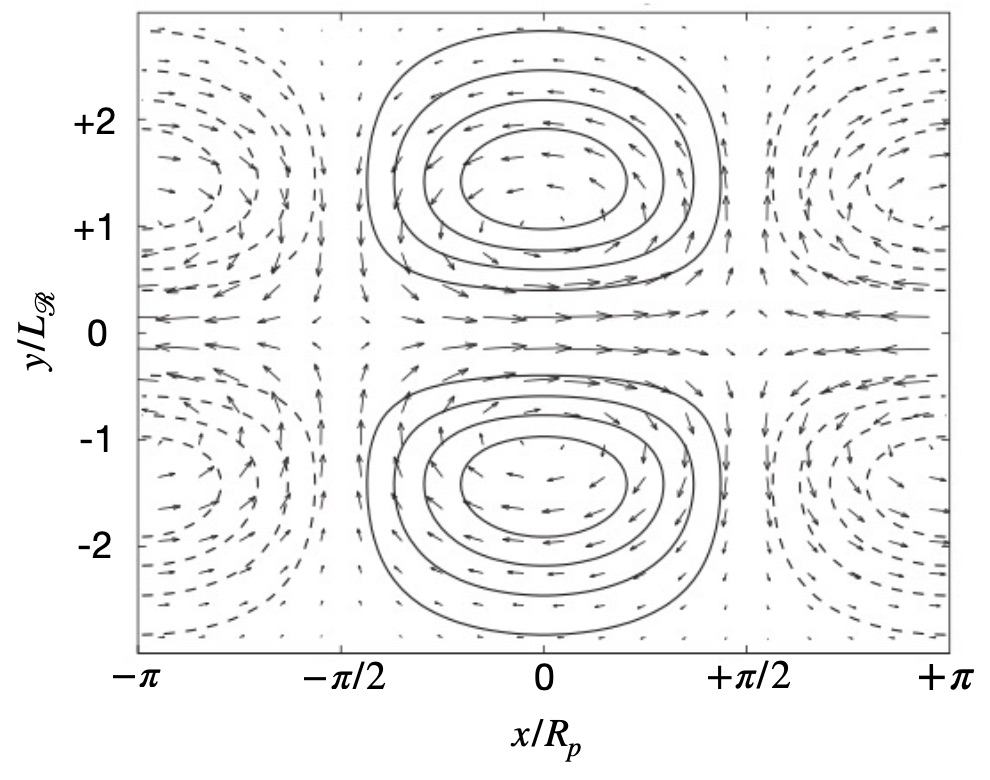}}
  \caption{Velocity (vectors) and height perturbation (contours) distributions
    for the equatorial Rossby wave with $(k,l) = (1,1)$, illustrating the
    horizontal structure of the dominant flow pattern in this paper (e.g. bottom
    $p$-level of Fig.~\ref{fig:heton}).  Note the different scalings in the $x$
    and $y$ directions, appropriate for the dynamics under discussion.  For the
    contours, negative (positive) values are full (dashed); under geostrophy,
    height and vorticity perturbation fields are locally (and at the equator)
    related by the Laplacian and opposite signs in spectral amplitude. }
 \label{fig:rwave}
\end{figure}

The velocity and height perturbation distributions for the equatorial Rossby
wave with $(k,l) = (1,1)$ from equation~(\ref{eq:disper}) are shown in
Fig.~\ref{fig:rwave}.  The figure illustrates the horizontal structure of the
dominant flow pattern in this paper, a modon pair.  We note that Rossby waves
generally have much lower phase velocities than inertia–gravity waves.  In
addition, in the long-wave part of the wave spectrum, there is a well-pronounced
gap between inertia-gravity waves and the rest of the spectrum.  In the next
section, we show how such a structure is modified in its morphology and
dynamical behaviour under full non-linearity and forcing at high resolution.

\section{Results}\label{results}
  
\subsection{Initial Structures} \label{sec:heton}

\begin{figure} %
  \centerline{\includegraphics[scale=.145]{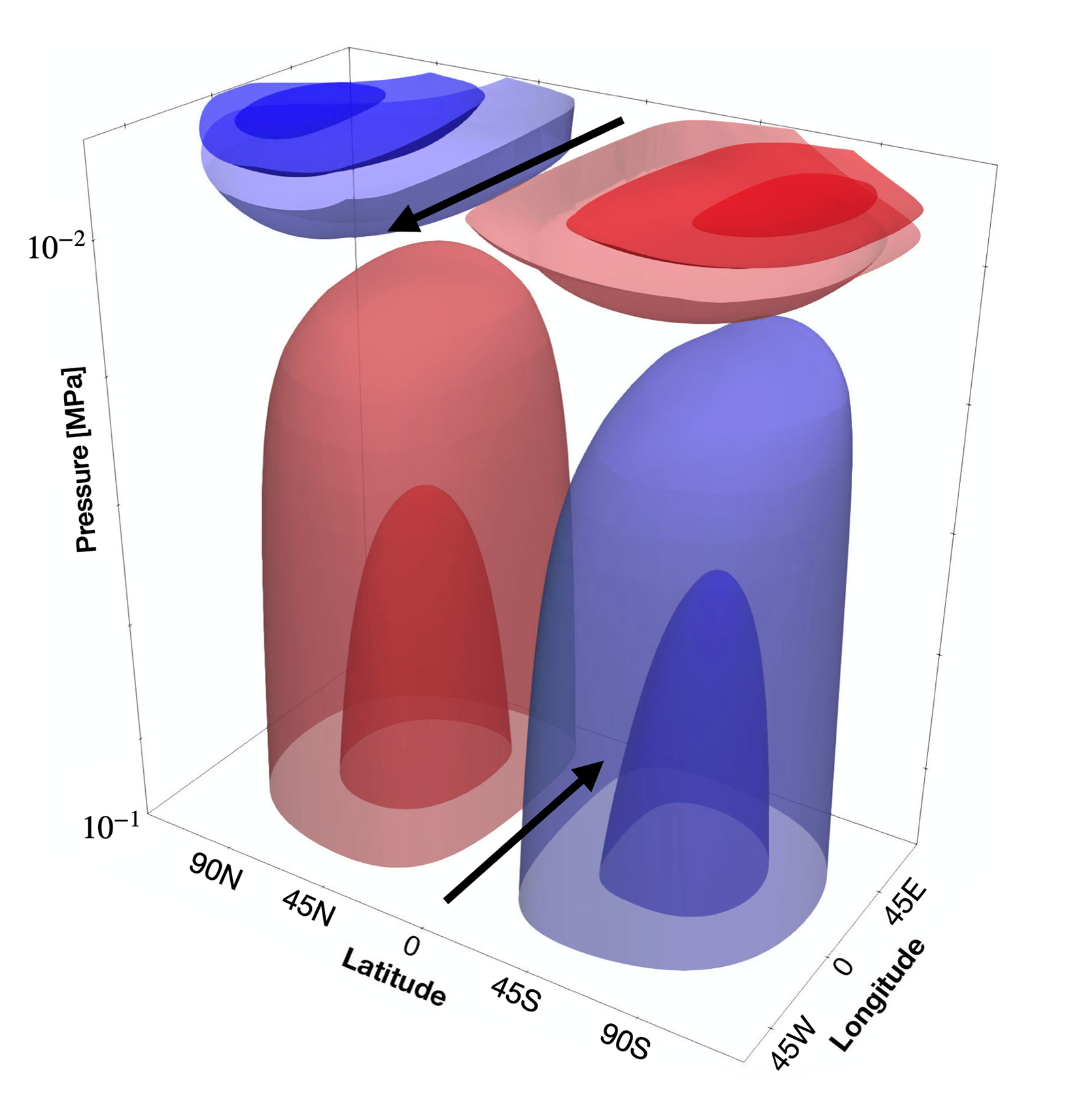}}
  \caption{Relative vorticity surfaces $\zeta(\lambda, \phi, p) = \pm 3$ (in
    units of $\tau^{-1}$ and at $\tau^{-1}$ intervals, with positive values in
    red and negative values in blue), at $t = 0.25$ (in units of $\tau$) from a
    T341L20 simulation: the dissipation order is $\mathfrak{p} = 8$, viscosity
    coefficient is $\nu_{16} = 1.5\times10^{-43}$ (in units of
    $R_p^{2\mathfrak{p}}\,\tau^{-1}$), and time-step size is $\Delta t =
    4\!\times\!  10^{-5}$ (in units of~$\tau$).  Black arrows show
      bulk velocity ($\sim\! 500$m/s). The section of the atmosphere shown is
    centred on the sub-stellar point $(\lambda, \phi) = (0,0)$ with vertical
    range $p \in [0, 0.1]$ (in units of MPa).  When the simulations are
    initialised as described in Section~\ref{setup} (i.e. baroclinic thermal
    forcing applied to atmosphere at rest), a ``hetonic quartet'' forms at the
    sub-stellar point; a second quartet forms at the night-side.  This
    baroclinic structure cannot be captured by a barotropic model
    (e.g. shallow-water model in Section~\ref{linear}). The quartet is composed
    of two oppositely-signed hetons, which here is anti-symmetric across the
    equator and separated in latitude by $\phi \sim 80^{\circ}$.  Each heton is
    a columnar structure composed of a pair of similar strength and
    oppositely-signed vortices at different $p$ levels (which here are tilted in
    longitude by $\lambda \sim 20^{\circ}$).  The heton quartet is a transient
    feature of the employed setup and is broken up by a strong vertical shear
    with the upper part of the atmosphere accelerating to a fast equatorial jet.
    The formation is independent of the location of the $p_{\rm bot}$ considered
    -- i.e. $p_{\rm bot} \in [0.1,10]$. }
 \label{fig:heton}
\end{figure}

Fig.~\ref{fig:heton} presents surfaces of relative vorticity ($\zeta\! =\! \pm
3$, in units of $\tau^{-1}$ and at $\tau^{-1}$ intervals), centred at the
sub-stellar point $(\lambda\! =\! 0,\, \phi\! =\! 0)$, at $t\! =\! 0.25$.  The
red and blue surfaces correspond to positive- and negative-valued surfaces,
respectively.  The overall structure seen in the figure is a pair of vortex
columns, each with opposite signs of $\zeta$ in the vertical direction. The
vortex columns form almost immediately after the start of the simulations and
then quickly (within a few $\tau$'s) evolve to a simpler structure, as described
below.  In oceanography, such columnar structures are known as ``hetons''
\citep{hoggstom85}.  Hetons are generated as a direct response to the applied
thermal forcing and potential vorticity \citep[e.g.][]{Pedl87} conservation,
which is valid for this setup at early times and wherever the flow is in
equilibration with the applied thermal forcing.  The snapshot shown is from a
T341L20 resolution simulation with a vertical domain range of $p \in [0, 0.1]$
and uniformly-spaced layers of $\Delta p = 5\!\times\!  10^{-3}$ thickness.

As seen in Fig.~\ref{fig:heton}, this early time flow structure is composed of a
{\it pair} of hetons -- i.e. a ``hetonic quartet'' \citep{kizner06}.  This
hetonic quartet is a vortical quadrupole which forms in response to the
specified thermal forcing.  The two hetons, as a pair, straddle the equator and
their centres are separated by a latitudinal distance of $\sim\!  80^\circ$.
Both hetons are slightly tilted in the longitudinal direction, with the top
vortical structure offset longitudinally by $\sim\!  20^\circ$ from the bottom
vortical structure.  Also, both hetons are cyclonic\footnote{Cyclonicity is
  defined according to the sign of $\bfzeta\cdot\bfOmega$: the sign is positive
  for cyclones and negative for anti-cyclones.} in the $p \la 2.5\!\times\!
10^{-2}$ region and anti-cyclonic in the $p \ga 2.5\!\times\!  10^{-2}$ region.
Note that the $p$-level where the vorticity inversion occurs depends on the
extent of the simulation's vertical domain (e.g. at a deeper level for larger
domain range).  Note also that, in addition to the heton quartet shown, a weaker
and oppositely-signed heton quartet forms near the anti-stellar point; hence,
there are actually two (anti-symmetric) hetonic quartets -- i.e. a ``hetonic
octet''.  The night-side hetons are larger in areal extent laterally; this is
due to the specified thermal forcing, which consists of relaxing the night-side
temperature field to a uniform distribution that is much cooler than the
day-side.  Note that the octet formation is independent of the domain $p$-range
considered in this work, from $[0,3\!\times\! 10^{-2}]$ to $[0,20]$.

\begin{figure*} %
  \centerline{\includegraphics[scale=.098]{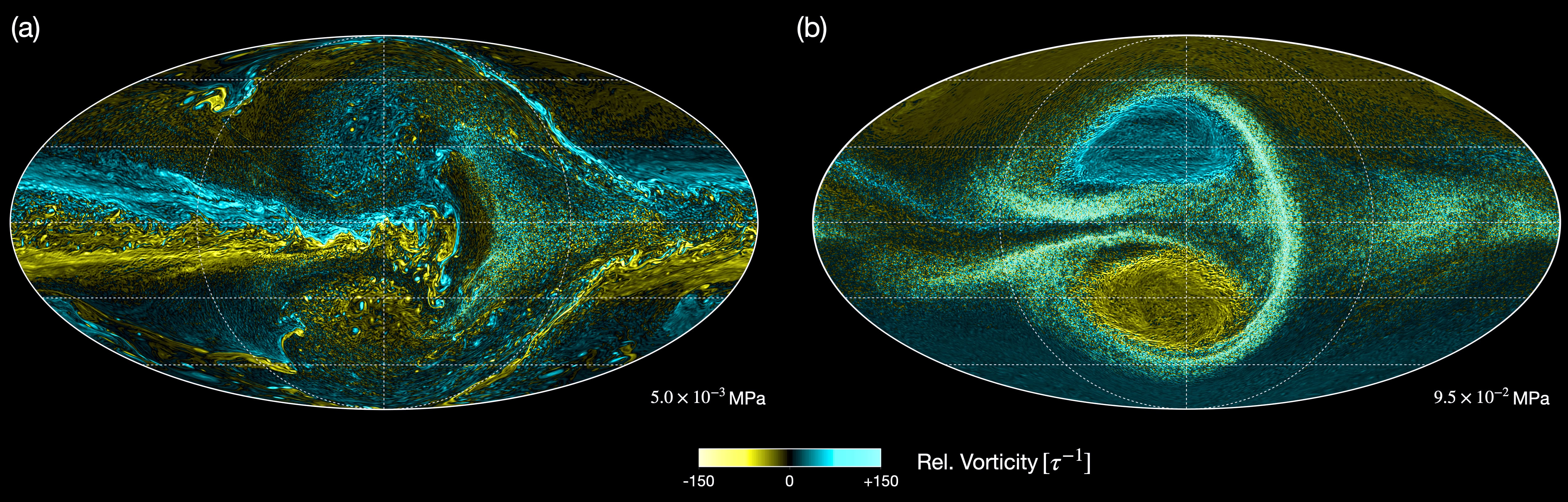}}
  \caption{The relative vorticity ($\zeta$) field from a T682L20 resolution
    simulation at $t\! =\!  100$ in Mollweide projection.  The dissipation
    order, viscosity coefficient and time-step size are $\mathfrak{p}\! =\!  8$,
    $\nu_{16} = 2.3\!\times\!  10^{-48}$ and $\Delta t = 2\!\times\!  10^{-5}$,
    respectively.  The $p$-levels shown, $5.0 \times 10^{-3}$~(a) and $9.5\times
    10^{-2}$ (b), correspond to the mid-points of the top and bottom layers of
    the computational domain, respectively.  The flow is dominated by two highly
    dynamic, planetary-scale modons -- a cyclonic modon at the sub-stellar point
    (at centres of the frames) and a much weaker and larger
    anti-cyclonic modon at the night side (the
    sides of the frames).  The cyclonic modon straddles an undulating,
    zonally-{\it a}symmetric equatorial jet at both $p$-levels.  The equatorial
    jet in a) is halted just to the east of the sub-stellar point and breaking
    throughout its ``core'' (along the equator): the jet core is where
    there is a jump in $\zeta$ in the meridional direction (near
    the equator).  The equatorial jet in b), in contrast, is rolling up much
    more prominently at its northern and southern edges.  The cyclonic modons in
    both frames emit large-amplitude gravity waves and generate thousands of
    small-scale vortices at their peripheries; the anticyclonic modons are
    barely visible at both $p$-levels because they are more diffused and lack
    sharp bounding fronts in these frames.  Both cyclonic and anticyclonic
    modons are highly dynamic and exhibit periodic life-cycles in which they
    generally (but not always) migrate westward around the planet, while
    strongly interacting with other flow structures -- e.g. storms, jets and
    waves.  }
 \label{fig:modon}
\end{figure*}

Overall, both quartets are strongly barotropic but quickly become baroclinic
(vertically tilted) as the thermal forcing accelerates the flow at the top of
the domain.  As early as $t\! =\! 1$, the night-side quartet migrates to the
day-side and the two quartets interact strongly with each other at the eastern
terminator.  This results in intense baroclinic fronts that sweep across the
eastern to western terminators, from the low latitude to the pole, and in both
the northern and southern hemispheres (the fronts, at a later time, can be seen
in Fig.~\ref{fig:modon}).  Significantly, the fronts undergo shear instability
and act as sources of small-scale vortices (storms): wind speeds parallel to the
fronts are already high, averaging $\sim$4.5~(i.e. $\sim$1500\,m\,s$^{-1}$) at
$p < 5 \times 10^{-2}$.  In general, {\it sharp fronts play a seminal role in
  hot exoplanet atmospheric dynamics}.  For example, capturing the front at the
early time (or any other time later) leads to a subsequent evolution which is
markedly different than when not captured.  The initially strong baroclinicity,
eventually shears off the top of the heton until like-signed vortical structures
vertically align; this leads to a strongly barotropic structure overall,
dominated by the modons which grew in strength upward from the bottom.  Thus,
hetons are transient structures: they may also form at later times but are
weaker and more short-lived, due to the ambient flow conditions (which is not at
rest).

\subsection{Modon Pairs and Other Storms}

Fig.~\ref{fig:modon} presents the main result of this paper: {\em given the
  idealised setup of a tidally synchronised exoplanet as described above, a pair
  of planetary-scale, strongly-barotropic modons is a generic solution to
  equation~(\ref{eq:pe})}.  Under the applied forcing, a pair of modons
(cyclonic and anti-cyclonic) forms near the planet's sub-stellar point
($\lambda\!  =\! 0$, $\phi\! =\! 0$) and the anti-stellar point ($\lambda\! =\!
180$, $\phi\!  =\! 0$).  Both modons straddle the equator, but the night-side
modon (anti-cyclonic) is always weaker than the day-side modon (cyclonic), when
they first form.  The cyclonic modon can be clearly seen in the figure, which
shows the $\zeta$-field at $t\! =\! 100$ -- well after the initial transient
behaviour (generally lasting less than $\sim$10 planetary days).  The resolution
of the simulation in the figure is T682L20; the dissipation order is
$\mathfrak{p} = 8$; the viscosity coefficient is $\nu_{16} = 2.3\!\times\!
10^{-48}$; the time-step size is $\Delta t = 2\!\times\!  10^{-5}$; and, the
extent of the vertical domain is $p \in [0,0.1]$.  The two levels presented,
which are from near the top and bottom of the simulation domain, show the
strongly barotropic structure of the modon.  Note that there is 
no $\zeta$ inversion between the $p$ levels shown in Fig.~\ref{fig:modon}.
We have performed an extensive
convergence assessment with the present setup and have verified that the
features and behaviours presented in this paper are qualitatively robust at, or
above, T341 horizontal resolution with $\nabla^{16}$ hyper-viscosity
\citep{SkinCho21}.

\begin{figure*} %
  \centerline{\includegraphics[scale=.205]{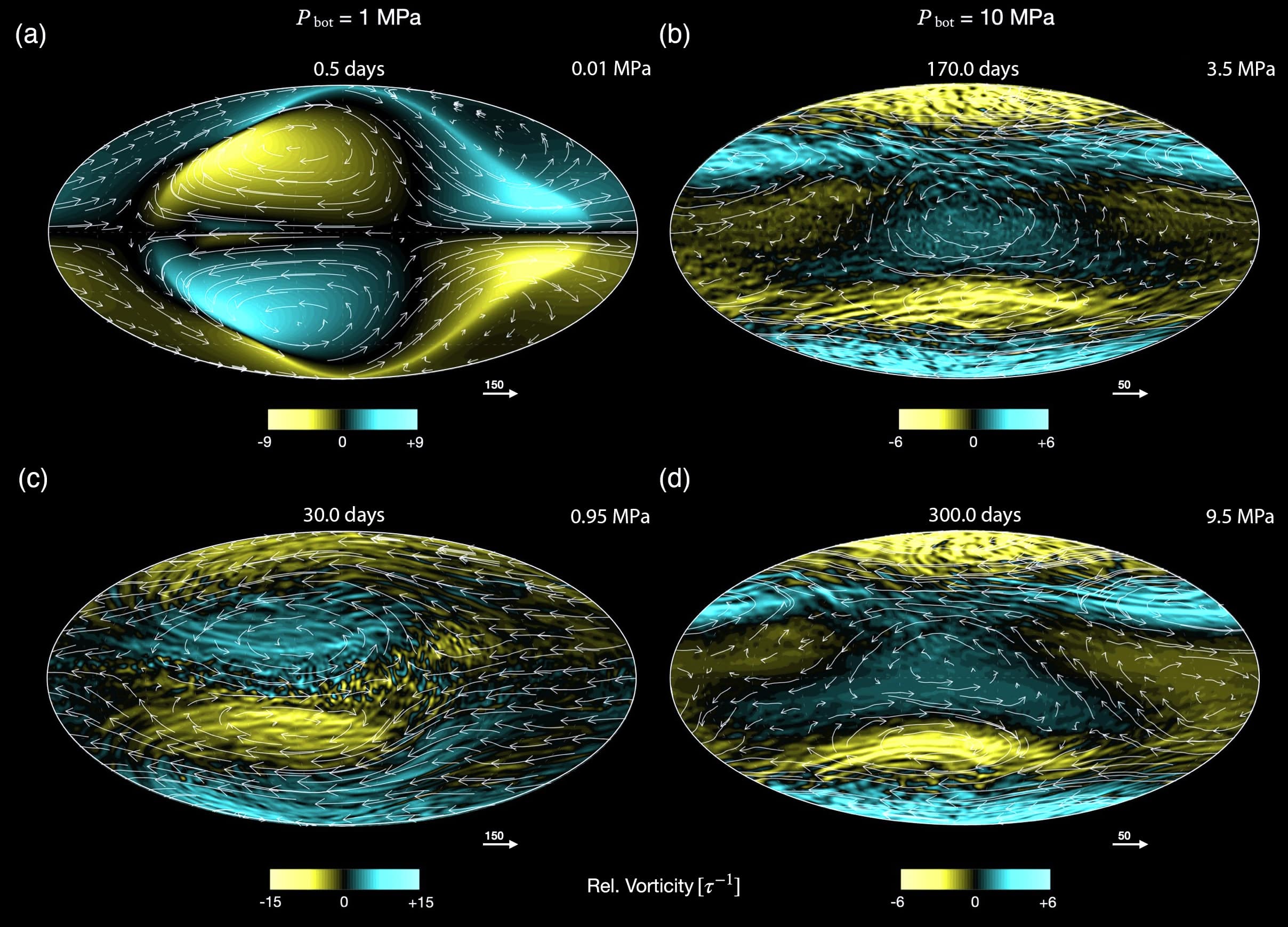}}
  \caption{Two ``deep atmosphere'' simulations with $p_{\rm bot} = 1$ (left) and
    $p_{\rm bot} = 10$ (right) at different times (labeled, top).
    The resolution of both simulations are T170L200; both are set up identically
    and carried out with $\mathfrak{p} = 8$, $\nu_{16} = 10^{-38}$ and $\Delta t
    = 8\!\times\!  10^{-5}$.  The $\zeta$ fields at the times indicated are
    shown in Mollweide projection centred on the sub-stellar point, and the
    times are chosen when the modons are clearly in view (near the sub-stellar
    point); the winds are overlaid with the reference vector length (in units of
    m\,s$^{-1}$, which is equal to $3.025\!\times\! 10^{-3}$ in $R_p\,\tau^{-1}$
    units), adjusted in each frame for visual clarity.  Frame~(a) shows the top
    of a {\it hetonic} quartet, which also forms in these deep
    simulations at early times (as in the ``shallow atmosphere'' simulations);
    the flow field is not yet turbulent at this time, but sharp fronts have
    already developed.  Frames~(b--d) show coherent modons are present at
    various times and at various $p$-levels.  Post the initial transient stage,
    the modons evolve, embedded in turbulent fields containing many smaller
    scale flow structures -- including gravity waves (striations in $\zeta$).
    Here modons generally migrate westward (chaotically) around the planet --
    periodically breaking or transitioning to different configuration states
    (again, as in the shallow atmosphere simulations).  Broadly, the flow tends
    to be more zonal and often contains an extra planetary-scale cyclonic modon
    as $p_{\rm bot} \rightarrow 20$, particularly in the very deep fields (b and
    d), compared with the flow in the shallow atmosphere simulations at this
    (and lower) horizontal resolution; but, the flow becomes much less zonal and
    is similar to that of the shallow atmosphere simulation at a higher
    horizontal resolution -- particularly in the region $p \la 1$
    \citep{SkinCho21,Choetal21}. }
 \label{fig:deep_modons}
\end{figure*}

\begin{figure*}%
  \vspace*{1cm}
  \centerline{\includegraphics[scale=.185]{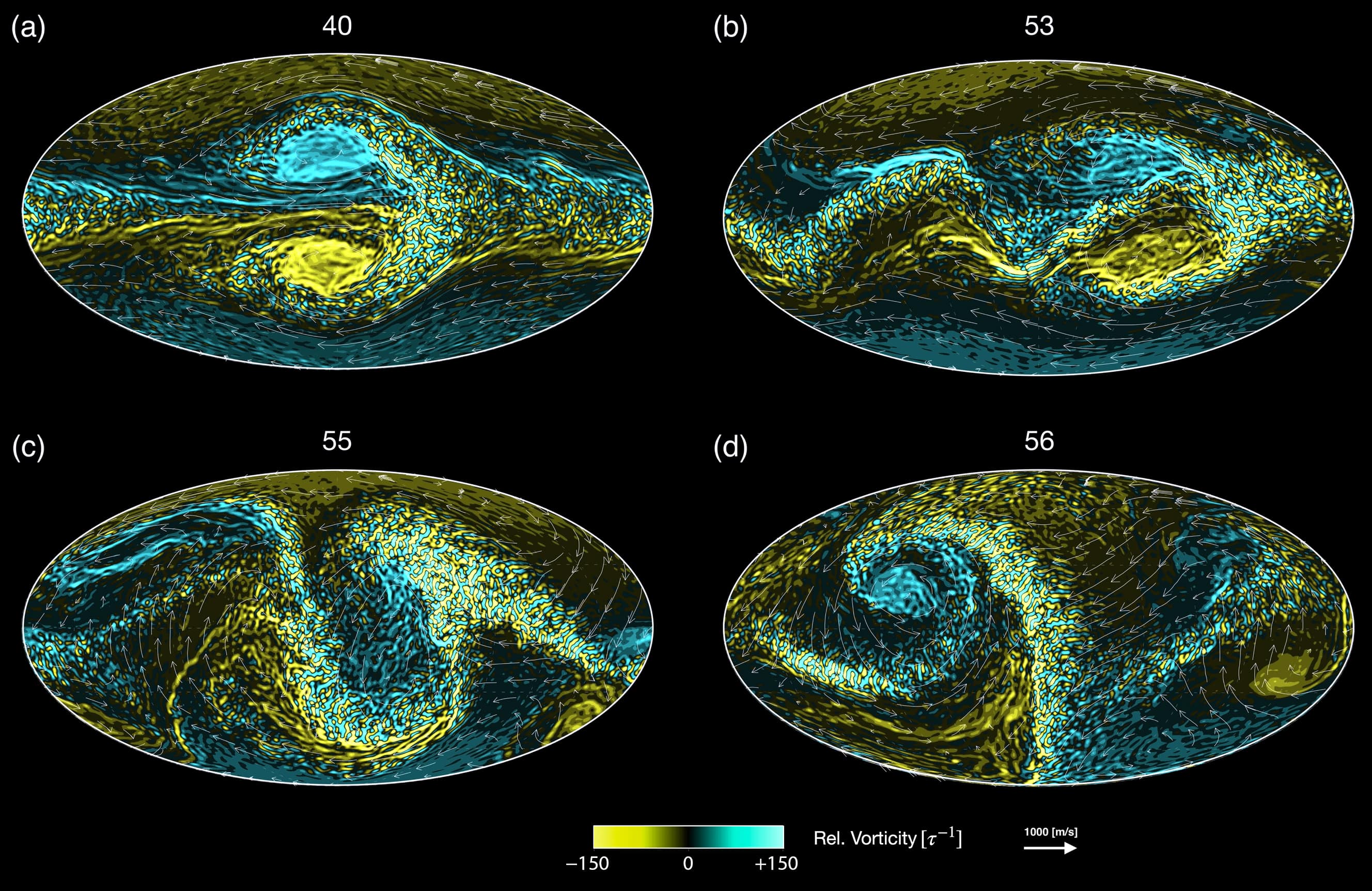}}
  \caption{ The $\zeta$ field, with overlaid winds (reference vector length is
    1000\,m\,s$^{-1}$, equal to 3.025 in $R_p\,\tau^{-1}$ units) at $p = 0.1$
    from a T341L20 simulation,\ with $\mathfrak{p} =8$, $\nu_{16} =
    1.5\times10^{-43}$, $\Delta t = 4\!\times\!  10^{-5}$ and vertical range of
    $[0, 0.1]$.  The four frames illustrate the onset of a secondary equilibrium
    state that arises in the simulations. In this sequence, a modon breaks apart
    into fast moving, uncoupled cyclones; and, the atmosphere is rapidly mixed
    as a consequence.  At $t = 40$ a cyclonic modon, which has spent an extended
    period of time positioned at the sub-stellar point, migrates eastward
    towards the night-side rather than continuing on its usual westward
    direction~(a).  As the modon reaches the eastern terminator ($t = 53$), it
    encounters fast flow from the night-side and shears apart, inducing very
    high amplitude undulations of the equatorial jet~(b).  At $t = 55$, both the
    equatorial jet and the cyclonic modon are converted into small-scale storms,
    which are transported westward around the planet by {\it very}
    high-amplitude Rossby waves~(c).  At $t = 56$, the small-scale storms
    coalesce into large-scale cyclones and the polar vortices break up~(d).  The
    atmosphere undergoes a period of vigorous mixing before a new modon forms on
    the day-side.  This sequence of patterns recurs quasi-periodically over the
    duration of the simulation. }
 \label{fig:breakup}
\end{figure*}

As can be seen in Fig.~\ref{fig:modon}, the equilibrated flow for this setup is
broadly characterised by three features: 1)~a fast, prograde (eastward flowing),
 zonally {\it a}symmetric equatorial jet; 2)~sharp,
planetary-scale fronts which form in both the northern and southern hemispheres
and roll up into small-scale vortices near the eastern terminator, particularly
at the lower $p$-level; and, 3)~planetary-scale storms that exhibit a variety of
quasi-periodic stable states, as well as transitions between those states.  We
stress that, when the flow is adequately resolved, the equatorial jet is {\it
  not} zonally symmetric; this is in contrast with nearly all simulations in the
past employing the same setup with lower resolution and/or viscosity order
\citep[e.g.][]{LiuShow13}.  Likewise for the unstable fronts; high-resolution is
required to capture this important source of medium-scale storms
\citep{Choetal03,Choetal21,SkinCho21}.  As for the planetary-scale storms, the
most generic state is a vortical quadrupole (at a given $p$-level), of which at
least one is a coherent modon (Fig.~\ref{fig:modon}, and see also
Figs.~\ref{fig:deep_modons} and~\ref{fig:breakup}).  In general, there are two
coherent modons, one composed of a pair of cyclones and the other composed of a
weaker pair of anticyclones, as already mentioned.  In the figure, the storms
comprising the cyclonic modon are separated meridionally by a centre-to-centre
distance of $\sim\!\pi R_p$: the storms of the anticyclonic modon are much less
conjoined and located closer to the poles.  Additionally, the cyclonic modon is
generally shrouded by a sharp front to its east and a trailing Rossby wave
mostly to its west (seen as undulations along fronts propogating
  westward from the cyclonic modon in Fig.~\ref{fig:modon}b), both of which
generate copious small-scale storms and gravity waves.  This is part of the
``geostrophic adjustment'' process of a modon \citep[e.g.][]{LahaZeit12}.  The
multiple flow states will be described separately in more detail later.

Broadly, the flow features described above are seen at essentially all the
$p$-levels.  Unlike the transient hetons described above, neither of the
cyclonic or anticyclonic modons in the figure changes its cyclonicity throughout
the simulation.  Modons are generally present near the top of the modelled
atmosphere (Fig.~\ref{fig:modon}a), although it is less prominent than at higher
$p$-level (Fig.~\ref{fig:modon}b): this is because of the stronger anisotropic
turbulence at the top.  Nevertheless, jets and fronts are both greatly
influenced by the modons, independent of $p$.  In particular, when the cyclonic
modon is positioned near the sub-stellar point the equatorial jet generally
steepens, undulates and then breaks on a time-scale of $\sim$3~planetary days.
This strongly non-linear behaviour is caused by the oscillation and rotation of
the storms (that comprise the cyclonic modon) about their equilibrium positions.
Note also how the two constituent storms have entrained small-scale storms from
the breaking jet into their cores (e.g. Fig~\ref{fig:modon}a),
thus mixing low-latitude air (and temperature) into the mid-latitudes.

Each modon forms out of an initial hetonic quartet, which strengthens upward in
time from the lower vortex after the tops of the original hetons are sheared
away (as described in Section~\ref{sec:heton}).  The first separation can occur
as early as $t = 5$, resulting in chaotic mixing in the top region of the
modelled atmosphere very quickly.  Despite the meridional symmetry imposed by
the initial and forcing conditions, the flow and mixing break this symmetry
early on.  The zonal symmetry is also broken early on, even before the zonally
asymmetric forcing builds up its strength (recall the short relaxation time
$\tau_{\rm th}$).  Both the modon and the front radiate large-amplitude gravity
waves as well as induce formation of small-scale storms.  The anti-cyclonic
modons, which are less visible in Fig.~\ref{fig:modon}, also radiate and
generate small-scale structures, but these structures are weaker and more
diffused.

After the initial formation, the modons execute a complex set of motions as they
evolve non-linearly.  The motions, however, exhibit distinct life-cycles, which
may be observable.  One of the cycles is associated with generation then
decimation of modons repeating many times over 1500 planetary days of T170
simulations.\footnote{Given the lack of realistic physical parameterizations,
  such a long duration simulation should not be taken too literally --
  particularly given the short $\tau_{\rm th}$ in the upper part of the domain.}
Qualitatively same behaviour occurs in T341 and T682 resolution simulation over
shorter durations (300 planetary days).  Significantly, both
cyclonic and anti-cyclonic modons are highly dynamic and chaotically translate
westward, in general, while undergoing frequent changes in size, shape,
orientation and strength as they do so.  These behaviours can easily be missed
when averaged fields or quantities are strictly used to study the atmosphere;
see e.g. discussions in \citet{Choetal15} and \citet{Choetal19}.  As already
discussed, both hetons and modons can form in simulations encompassing a wide
range of pressure levels.  This includes simulations which include the deeper
region, where the thermal forcing is not applied.  This is shown explicitly in
Fig.~\ref{fig:deep_modons}.

In Fig.~\ref{fig:deep_modons}, two ``deep atmosphere'' simulations (i.e. those
with $p_{\rm bot} \ge 1$) are presented.  The resolution of the simulations is
T170L200; and, $\mathfrak{p} = 8$, $\nu_{16} = 10^{-38}$ and $\Delta t =
8\!\times\!  10^{-5}$.  The $p_{\rm bot}$ is different between the two
simulations, with values of 1 and 10 (left and right columns, respectively).
The $\zeta$-field is displayed in all the frames.  Fig.~\ref{fig:deep_modons}a
shows the top of a hetonic quartet, which forms in the deep atmosphere
simulations at early times -- as in the ``shallow atmosphere'' simulation
(i.e. with $p_\text{bot} = 0.1$, presented in Fig.~\ref{fig:heton}).  The
meridional symmetry is strong at the time shown, but quickly breaks shortly
thereafter.  Figs.~\ref{fig:deep_modons}(b--d) present modons in deep atmosphere
simulations at different $p$-levels and times.  As can be seen, modons occur
near the top of the modelled domain independently of the location of the domain
bottom, provided a sufficient number of vertical levels are used to span the
domain.  The modons shown here also translate around the planet and periodically
break and reform into several different configurations of vortices, similar to
the behaviour in the shallow atmosphere simulations.

We note here that several past studies employing a similar setup have captured
vortex dipole structures \citep[e.g.][]{ThraCho10,Hengetal11, Choetal15,
  SkinCho21}.  However modons were not the focus of these studies.  Moreover,
past studies were performed with lower resolution and/or stronger numerical
viscosity than in the present study.  Since capturing the fine-scale structures
(and, in particular, their influence on the large-scale structures) is crucial,
high-resolution and minimal over-dissipation are necessary ingredients in
accurately modelling modons.  Specifically, having extensively investigated the
influence of well-resolved fine-scale structures, we find the longevity,
dynamism and multiple state behaviours of planetary-scale modons are
fundamentally affected.  Below the T170 horizontal resolution and when low-order
viscosity and/or large viscosity coefficient are used, modons are markedly
diffused and translate smoothly around the planet -- {\it if} they move at all.
Strict numerical convergence does not appear to be achieved until at least T341
resolution \citep{SkinCho21}.

\begin{figure*} %
\vspace*{.8cm}
  \centerline{\includegraphics[scale=.089]{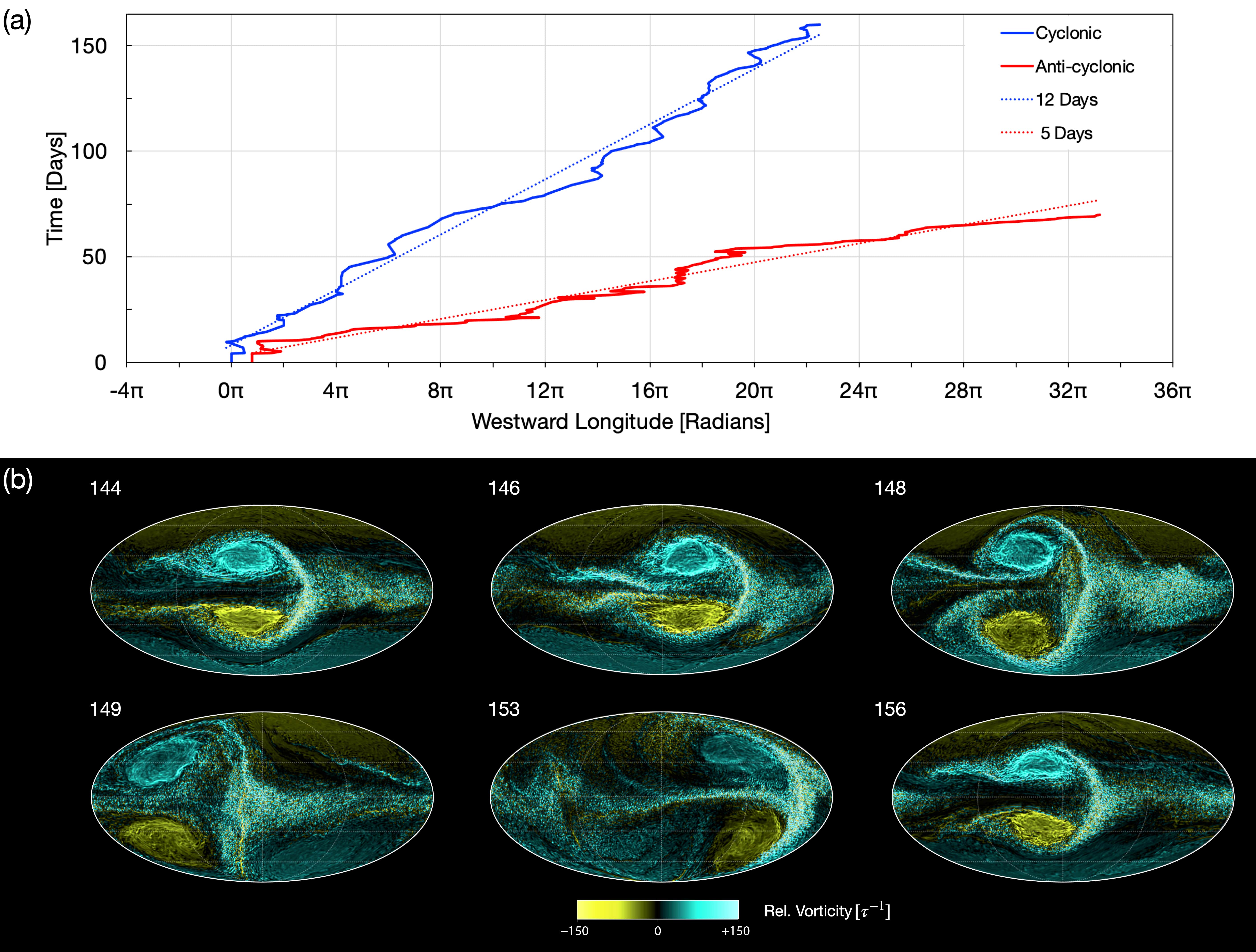}}
  \caption{(a) Longitude position of the modon centroids (in radians) westward of
    the sub-stellar point, as a function of time $t = [0, 125]$ and at $p =
    0.1$, from the simulation in Fig.~\ref{fig:modon}.  Blue and red lines
    correspond to the cyclonic and anticyclonic modon centroid trajectories,
    respectively.  Dotted lines are constant westward migration rate fits with
    periods 12 and 5, as indicated in the legend.  Both modons exhibit
    quasi-periodic life-cycles in which they translate chaotically westward
    around the planet in the bulk, undergoing multiple excursions about a smooth
    straight-line trajectory.  On multiple occasions, modons pause at the
    day-side (night-side) and oscillate to either side of the sub-stellar
    (anti-stellar) point, before continuing on.  Such behaviours are significant
    for observations. (b) The relative vorticity ($\zeta$) field from the simulation
    shown in Fig.~\ref{fig:modon} at $p\sim 0.1$ and in Mollweide projection showing 
    a single cycle of a modon's westward translation around the planet. }
  \label{fig:modon_track}
\end{figure*}

Fig.~\ref{fig:modon_track} presents the longitude positions of the modon
centroids from the T682L20 simulation in Fig.~\ref{fig:modon}, as an
illustrative example.  Such plots are instructive, particularly for
observations, because of the modons' strong influence in redistributing large
patches of hot and cold air across the planet -- as discussed more in detail
below.  In the figure, blue and red lines show the longitudinal distance
traversed by the cyclonic and anticyclonic modons, respectively.  Dotted lines
indicate linear fits, with the $2\pi$-traversal periods indicated in the legend.
As already noted, the behaviours are quantitatively different at different
resolution and dissipation order.  However, the general behaviour here is very
roughly similar down to T85 resolution -- provided $\mathfrak{p} = 8$
dissipation order is employed \citep{SkinCho21}: in particular, the cyclonic
modon's westward translation is observed, but the modon is sluggish and is
devoid of the non-linear or oscillatory motions (which arise when small-scale
eddies {\it are} captured at T341 and above resolutions).  Essentially all
previous extrasolar planet simulations have been performed with
less than T341 resolution and $\mathfrak{p} = 8$ dissipation order -- especially
ones that use the current setup and are three dimensional.

In the figure, modons are seen executing westward translations around the planet
in the bulk, on a few to $\sim$20 planet-day periods.  But, there are pauses,
reversals of direction and short period ($\la$~day) oscillations superimposed on
the bulk motion.  This is due to the strong interaction with small-scale
structures.  Frequently, the modons break and reform at the end of their
``life-cycles'' (e.g. Fig.~\ref{fig:breakup}).  However, while both types of
modon feature quasi-periodic life-cycles, they generally display large
differences in periodicity and phase between them -- after the initial,
formative period of $\sim$10~planetary days.

In a typical life-cycle of the cyclonic modon, the modon first forms slightly
westward of the sub-stellar point and initially translates eastward --
consistent with its cyclonic character {\it in the background of no, or very
  weak, motion}.  After reaching the sub-stellar point, the modon often
``hangs'' in this position for an extended period of time (up to
$\sim$10~ planetary days).\footnote{In some cases, it can be as
  long as $\sim$100~planetary days initially, if the zonal
  symmetry remains unbroken or only very mildly broken (e.g. when the forcing is
  more symmetric or dissipation steers the flow).}  Near the sub-stellar point,
the cyclonic modon generates and interacts with a very large number of
small-scale storms of opposing signs to the cyclonic modon in the northern and
southern hemispheres (Fig.~\ref{fig:modon}b).  This effectuates the meridional
(as well as enhances the zonal) symmetry breaking, causing the modon to
oscillate about the sub-stellar point by $\sim$10$^\circ$ in longitude.  This is
concomitant with a decrease in the modon's size and strength as well as an
increase in angular separation between the constituent cyclones, thus rendering
the modon susceptible to further perturbations.  This is a characteristic
property of modons in the ageostrophic regime.  After the extended period near
the sub-stellar point, the modon moves westward, past the western terminator and
quickly traverses the night-side to reach the eastern terminator in a nearly
continuous motion.  The traversal usually takes no longer than
$\sim$5~planetary days.  Finally, at the eastern terminator, the
modon's constituent cyclones uncouple and dissipate completely or move to high
latitudes -- whence the cycle ends.  A new cyclonic modon subsequently forms on
the day-side and the cycle begins again.  In total, the cycle lasts for
$\sim$15\,($\pm 4$)~planetary days, of which the modon generally
occupies the day-side for $\sim$10~planetary days.

While the anticyclonic modon's life-cycle is qualitatively similar to that the
of the cyclonic modon, it is shifted in phase (longitudinal location) and has a
considerably shorter migration period.  The latter is because the anti-cyclonic
modon is generally located nearer to the planet's poles and precesses only
slightly off the polar axis at high-latitudes, rather than traversing the entire
circumference of the planet near the equator.  Ultimately, this is related to
the asymmetry of the thermal forcing (recall that the night-side equilibrium
temperature is uniform and lower than the day-side temperature).  Consequently,
the variability can be considerably different, depending on the latitude being
viewed.  This points to the importance of knowing the planet's obliquity (or
inclination angle), when interpreting atmospheric variability.  Further, note
that the cyclonic and anti-cyclonic modon tracks are roughly anti-correlated in
longitude in Fig.~\ref{fig:modon_track}; for example, the anti-cyclonic modon
forms on the night-side at the same time as the cyclonic modon, which forms on
the day-side.  Such coupled behaviour presents additional ``targets'' for
current and future observations.

\newpage 
The above discussion suggests a multitude of quasi-equilibrium flow states \footnote{We define ``quasi-equilibrium'' 
as a state that persists over a long duration -- i.e. over a period much longer than $\max \{\tau_{\rm th}, \tau_{\rm ad}\}$, 
when $\tau_{\rm th}<\infty$.}. These states are associated 
with distinct flow configurations and periodicities
and lasts for extended periods of varying durations.  Moreover, multiple
transitions between the states within a single simulation occur as well.  In
general, the set of available states vary with $p$-levels; but, they are often
correlated.  The overall behaviour is generic across small variations in thermal
forcing profile and initial flow condition.  However, as expected, the behaviours
is {\it quantitatively} different at lower resolution and even {\it
  qualitatively} different below T85 resolution.  The latter is because
simulations below the T85 resolution (particularly with $\frakp < 8$) produce
modons (when they are able) that are not dynamic \citep{SkinCho21}.

Fig.~\ref{fig:breakup} presents an example of a state transition that frequently
occurs in the current setup.  The four frames in the figure illustrate the key
stages of a transition between two states known as ``blocking patterns'' -- a
``diffluent block'' and an ``omega block''.\footnote{In meteorology, blocks are
  large-scale weather patterns that are nearly stationary and effectively block
  and redirect flows \citep{Rex50,Wooletal18}.}  In the figure, blocking actions
rapidly mix and chaotically churn the atmosphere over a several-day time-scale.
The transition is from the diffluent block to an omega block and eventually back
to the diffluent block (often after a complete dispersal of the original modon,
as shown).  We note that the modon in this example had previously completed many
traversals of the planet prior to this breakup.  Instead of continuing its
westward translation, this modon translates eastward across the sub-stellar
point close to the eastern terminator (Fig.~\ref{fig:breakup}a), where it starts
to tilt in the counter-clockwise direction (Fig.~\ref{fig:breakup}b).  In the
latter, the shearing modon disrupts the equatorial jet, generating many hundreds
of small-scale storms in the process.  After one to two planetary days, the
modon is nearly sideways, fully into the omega state (Fig.~\ref{fig:breakup}c).
At this point, the small-scale storms generated are carried westward from the
night-side to the day-side by ({\it very}) high-amplitude Rossby waves
(e.g. undulation of the equatorial jet).  Fig.~\ref{fig:breakup}d shows the
resulting flow pattern after the small scale-storms have coalesced into a large,
energetic cyclone that veers off northwest, into the night-side.  Unlike the
modons, these uncoupled cyclones rapidly traverse the planet, inducing further
vigorous mixing.  The whole process causes the atmosphere to be much more
homogenised.  Note that studies have demonstrated that higher resolution
improves the representation of the blocking in climate models (of the Earth),
when compared with observations \citep[e.g.][]{Anstetal13}.  Thus, we reiterate
here one of our major points in this paper: {\it higher resolution, than
  employed in the past, is necessary to capture the atmospheric states of hot
  extrasolar planets accurately}.

\subsection{Dynamic Temperature Redistribution}

In this subsection, we discuss more directly the effects of modons on the
temperature distribution of the atmosphere -- and, more broadly, observable
signatures produced by them.  Our discussion here centres mainly on the shallow
atmosphere simulations; however, we also present and discuss deep atmosphere
simulations.  One reason for the shallow atmosphere focus is that there is
little qualitative difference between simulations with $p_{\rm bot} \in \{0.1,
1.0\}$ (recall that the applied thermal forcing does not extend to $p \ge 1$
region).  Another important reason is that adequately resolving -- and choosing
-- the vertical range of the atmosphere (including choosing the location of
$p_{\rm bot}$) is still uncertain and computationally challenging in $p_{\rm
  bot} \gg 1$ cases \citep{SkinCho21}. Qualitatively, the
general flow pattern changes monotonically as $p_{\rm bot}$ increases from 0.1
to 1.0; and, the modons become weaker and wider with the equatorial jet becoming
more zonal, as $p_{\rm bot} \rightarrow 1$.  We have verified
that this general picture does not change qualitatively as $p_{\rm bot}
\rightarrow 20$, with up to T341L200 resolution.  Deep atmosphere simulations
are discussed below.  Note that the qualitatively robust general behaviour of
the modons is due to the strongly barotropic quality of the flow.  This is
intimately related to the setup employed.

\begin{figure}%
  \centerline{\includegraphics[scale=.11]{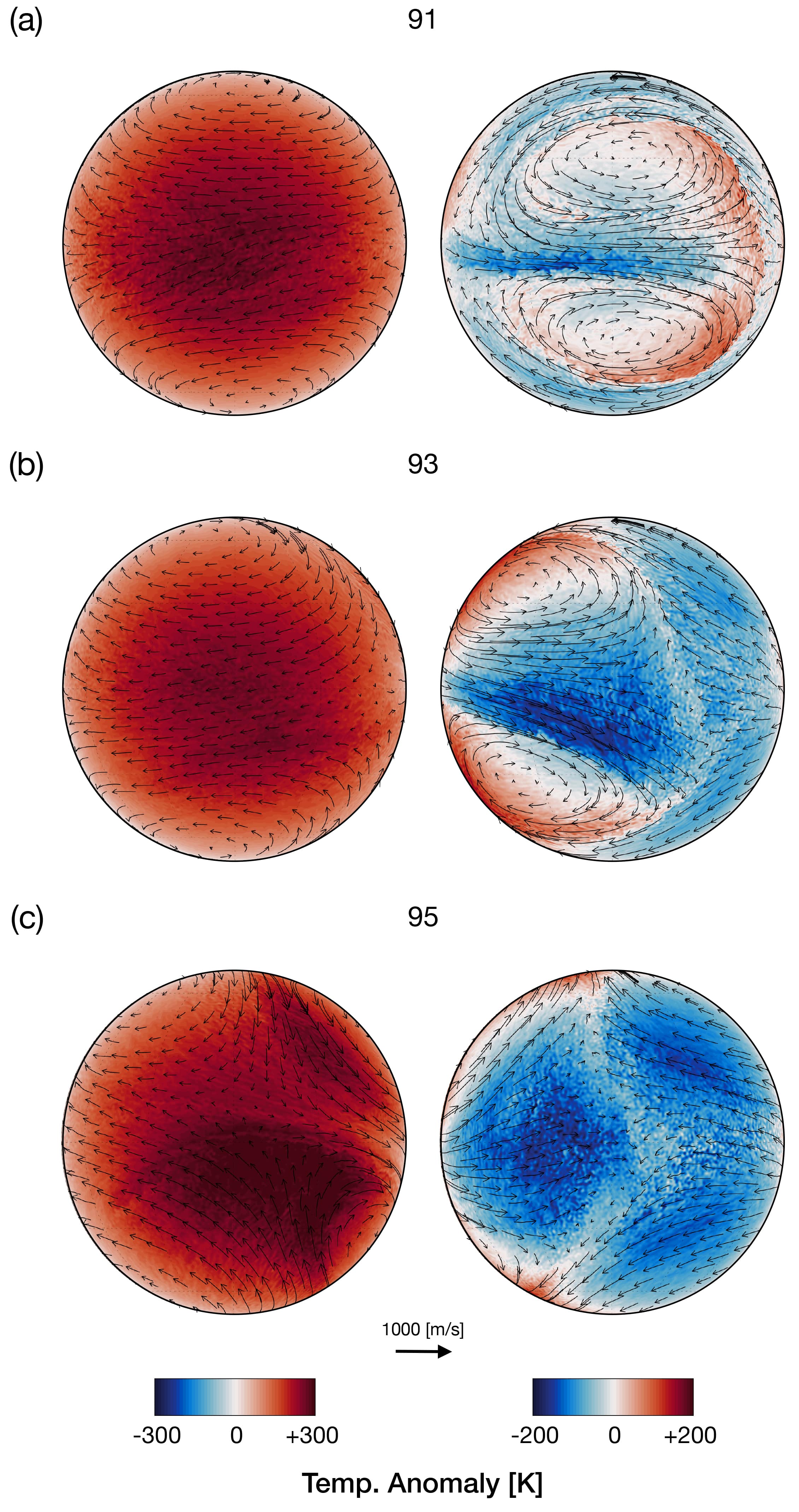}}
  \caption{Temperature anomaly field, $T_a(\lambda,\phi,p) = T - T_{\text{eq}}$,
    with the winds overlaid: fields from the simulation in
    Fig.~\ref{fig:breakup} are shown at $p = 5 \times 10^{-3}$ at the indicated
    times (however, the modons are in their ``quiescent'' cycle, in contrast
    that in Fig.~\ref{fig:breakup}).  Left and right columns show the
    night-side~(NS) and day-side (DS) discs, respectively; and, the projection
    is orthographic.  Red and blue colours indicate regions that are cooling ($T
    > T_\text{eq}$) and heating ($T < T_\text{eq}$), respectively: hence, they
    indicate regions which are absorbing and emitting, respectively.
    (a)~Day-side emission is localised in a narrow arc at the cyclonic modon's
    eastern periphery; the modon's two cyclones sequester hot air in their
    cores, but they are not emitting ($T_\text{eq} \approx 0$), due to the short
    relaxation time.  The day-side equatorial jet is strongly absorbing at this
    time and does not ``transport heat'', but rather cold temperature (see also
    $t =93$, DS).  The anti-stellar point is generally emitting because the
    equatorial jet and polar modon directs hot flow from the day-side.  (b)~When
    the cyclonic modon migrates to the western terminator, hot air from the
    day-side is transported to the night-side, around the periphery of the
    cyclonic modon via the poles.  (c)~A local patch of strong emission on the
    night-side ($T_a \approx 100$\,K) is generated.  Since the modons migrate
    periodically around the planet, the simulations exhibit periodic peaks in
    emission on the day {\it and} night sides.  }
  \label{fig:temp_anom}
\end{figure}

Fig.~\ref{fig:temp_anom} shows the instantaneous temperature anomaly field,
$T_a(\lambda,\phi) = T - T_{\text{eq}}$, at $p = 5\!\times\! 10^{-3}$.  The
$T_a$-fields are chosen from the T341L20 simulation shown in
Fig.~\ref{fig:breakup}, when the modons are not in one of the ``breakup''
cycles: the fields in a ``quiescent'' cycle are presented.  The fields at $p =
0.1$, corresponding to the bottom of the shallow atmosphere simulation, match
very closely with the presented fields because of the strong barotropic
character of the flow.  The $T_a$-field field reveals the regions of heating
($T_a < 0$) and cooling ($T_a > 0$), or absorbing and emitting, respectively.
For greater lucidity, the fields are shown in orthographic projection, centred
on the anti-stellar and sub-stellar points, illustrating the night-side (NS) and
day-side~(DS) discs separately.  Note that, in the absence of dynamics, $T_a =
0$ globally.  A primary aim of this work is to demonstrate clearly that the flow
induces a significant deviation from this -- even when $\tau_{\rm th} \ll
\tau_s$ (Fig.~\ref{fig:prof1}).  This is in contrast to what is reported in many
previous studies \citep[e.g.][]{Cooper05,Showmanetal08a,LiuShow13}.  Another major
aim is to clearly demonstrate that heat is transported mainly by eddies (storms)
rather than the equatorial jet, as is often asserted.

As can be seen in Fig.~\ref{fig:temp_anom}, the local emission and absorption
are strongly dependent on the location, size and strength of both modons.  For
example, at $t = 91$ on the day-side, temperature is localised to two ``hot
spots'' of $T \approx T_\text{eq}$, where each of the ``spot'' is really one of
the modon's cyclones sequestering hot air in its (extended) core
(Fig.~\ref{fig:temp_anom}a).  Importantly, the sequestered regions are neither
emitting nor absorbing thermal radiation since $T_a \approx 0$; hence, the
dynamics is also thermally unforced in these large regions.  Both the limb and
equatorial jet of the day-side disc are strongly absorbing, where $T_a \approx
-125$\,K.  In contrast, the night-side is emitting in a wide area near the
anti-stellar point with $T_a \approx +40$\,K.  At $t = 93$, as the modon
migrates from the sub-stellar point to the western terminator, hot air is
transported to the night-side towards the higher latitudes
(Fig.~\ref{fig:temp_anom}b).  Consequently, the day-side and night-side are now
in absorption and emission, respectively.  At $t = 95$, as the cyclonic modon
emerges on the night-side, the {\it westward} jet between the anti-cyclonic
modon begins to intensify at the equator and the meridional flow on the western
flanks of the cyclones converges to the anti-stellar point, transporting hot air
over the high latitudes from the day-side to the night side
(Fig.~\ref{fig:temp_anom}c).  This results in a more continuously heated
(i.e. emitting) night-side, with periodic bursts of strong emission.  The period
of oscillations (of both emission and absorption) is about 10 to 15 planetary
days, as described above (Fig.~\ref{fig:modon_track}).

\begin{figure*}%
  \vspace*{1cm}
  \centerline{\includegraphics[scale=.12]{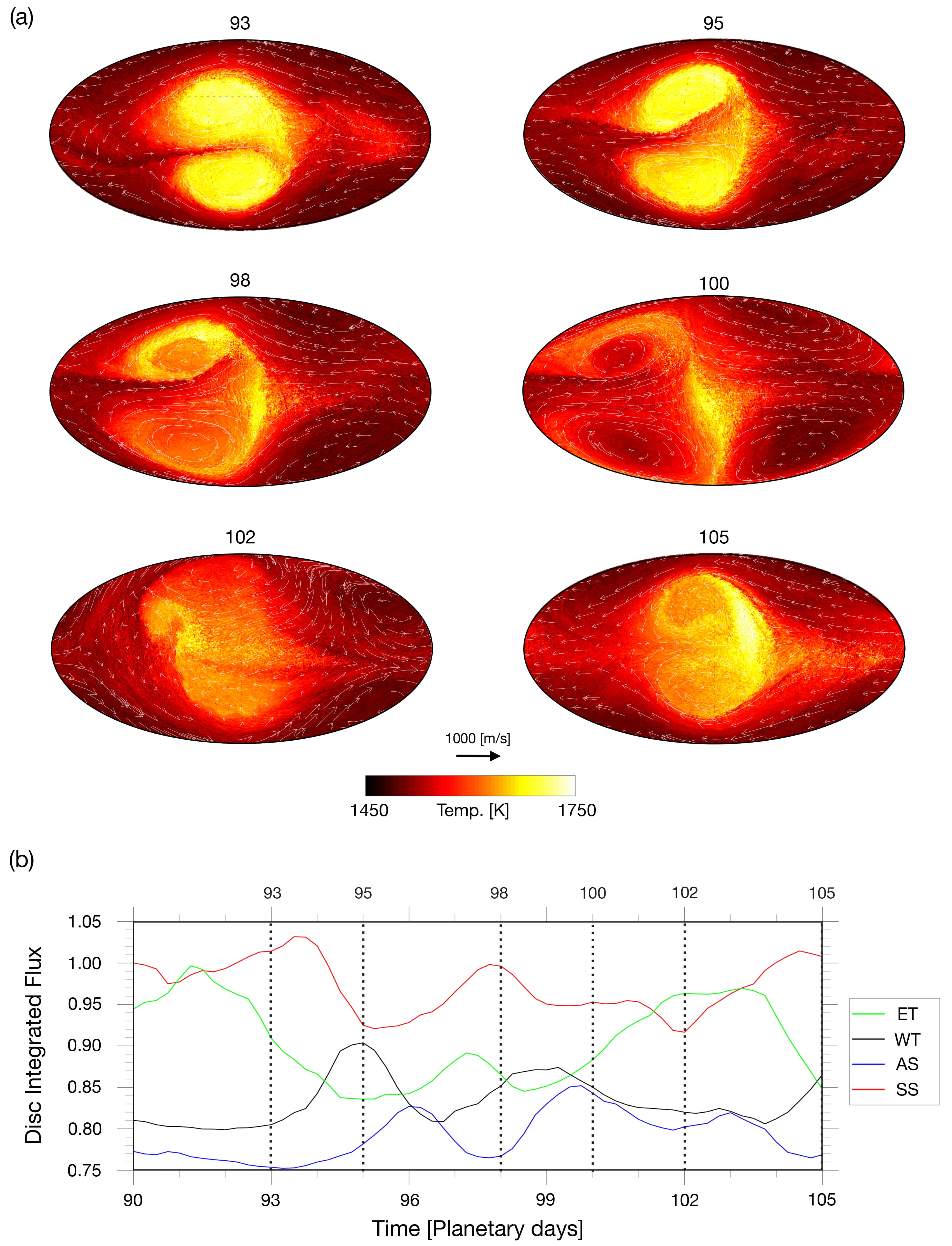}}
  \caption{The cyclonic modon's periodic migration zonally redistributes
    planetary-scale regions of hot (and cold) flow with $\sim$12 planetary day
    periodicity, and latitudinally mixes the temperature distribution of the
    atmosphere. (a) Instantaneous temperature fields (in units of $K$) with
    overlaid winds at the $p = 0.1$ level.  Time frames $t = [93, 105]$
    illustrate key stages in the life-cycle of a cyclonic modon and the
    corresponding global temperature distribution.  At $t = 93$, a modon forms
    near the sub-stellar point and migrates westward around the planet over a
    $12$-day period, in which it mixes and redistributes hot (cold) regions of
    the flow.  At $t = 102$, the cyclonic modon dissipates at the end of its
    life-cycle and subsequently reforms at $t = 103$.  The modon's effect on the
    global temperature distribution produces discernible signatures in the
    time-series of normalised disc-integrated flux, $T/T_{\rm
        ref}$~(b) at each of the planet's key locations: eastern terminator
    (ET), western terminator (WT), anti-stellar point (AS), and sub-stellar
    point (SS).  In general, the SS flux is the largest and the AS flux is the
    smallest of the four fluxes -- as expected, given the short $\tau_{\rm th}$.
    However, the variations in each flux is such that SS flux can be lower than
    the ET flux and the ET flux can be lower than the WT flux, at different
    times and with different periods.  Both (a) and (b) show that the hottest
    (coldest) region can be located over a broad range of longitude and on
    either side of the sub-stellar (anti-stellar) point.  }
  \label{fig:temp_flux}
\end{figure*}

Fig.~\ref{fig:temp_flux} shows the actual redistribution of temperature $T$ by
the cyclonic modon over the full traversal around the planet.
Fig.~\ref{fig:temp_flux}a shows the instantaneous temperature fields with
overlaid wind vectors in Mollweide projection at $p = 0.1$.  Six frames in the
interval, $t = [93, 105]$, illustrate various stages in this particular type of
{\it thermal} cycle.  We reiterate that, in the absence of dynamics, the $T$
field shown would resemble a simple circular patch of ``hot spot'' centered at
the sub-stellar point, instead of the two disjointed patches associated with the
modon's cyclones (as was seen in Fig.~\ref{fig:temp_anom}).  In
Fig.~\ref{fig:temp_flux}b, the disc-integrated flux time-series 
  (also at $p = 0.1$), encompassing the frames in Fig.~\ref{fig:temp_flux}a,
are presented for four of the planet's key locations: sub-stellar point (SS),
anti-stellar point (AS), eastern terminator (ET) and western terminator (WT).
At each of these locations, the local ``radiative equilibrium flux''
(i.e. $\sigma T^4$) from Fig.~\ref{fig:temp_flux}a is averaged
over the disc, weighted by a cosine factor to account for the spherical
geometry.\footnote{From hereon, we shall simply refer to this quantity as ``temperature flux'', 
or often just ``flux'', as in \citet{SkinCho21} and
  \citet{Choetal21}.}  The flux is normalised by the the initial flux based on
$T_{\rm ref}(p)$.  In the absence of additional physical parameterisations
(e.g. radiative transfer, cloud, and species distributions) which can be added
for increased physical realism, $\sigma T^4$ is an adequate measure of the
equilibrium flux.  A more realistic flux using a radiative transfer model will
be reported elsewhere.

\begin{figure*}%
  \vspace*{1cm}  
  \centerline{\includegraphics[scale=.305]{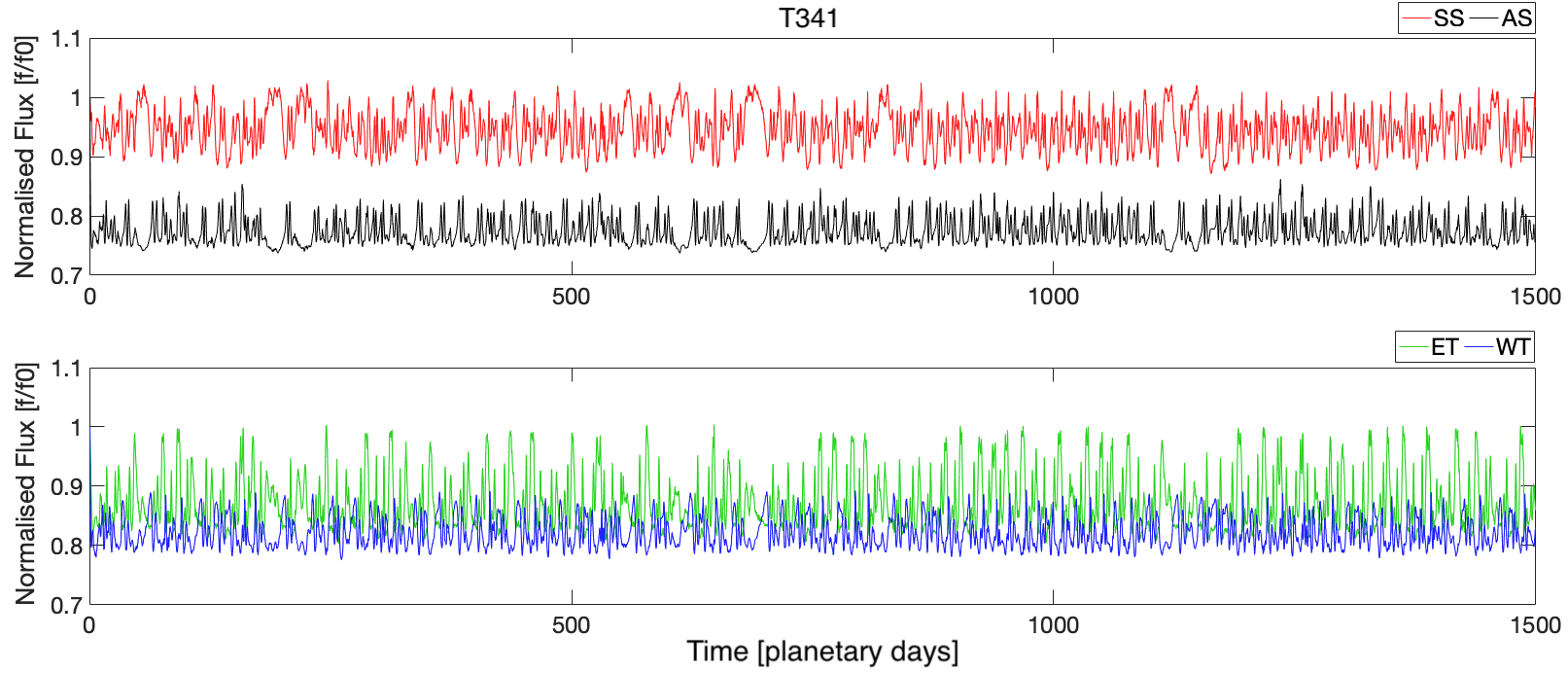}}
  \caption{Long duration time-series of normalised disc-integrated flux at $p =
    0.1$ for a simulation at T341L20 resolution with $\mathfrak{p} = 8$,
    $\nu_{16} = 1.5\times10^{-43}$ and $\Delta t = 4 \times 10^{-5}$.  Top and
    bottom panels show normalised disc-integrated flux at the sub-stellar (SS)
    and anti-stellar (AS) points and the eastern terminator~(ET) and western
    terminator (WT), respectively.  Quasi-periodic variability associated with
    the modon's behaviour (e.g. Fig.~\ref{fig:temp_flux}) persists for $1500$
    planetary days; hence, the simulations and behaviours presented above are
    assumed to be equilibrated over the duration shown.  The time-series
    behaviours can be directly related to temperature redistribution by the
    modons.  }
  \vspace*{.5cm}
  \label{fig:flux_long}
\end{figure*}

\begin{figure*}%
  \centerline{\includegraphics[scale=.22]{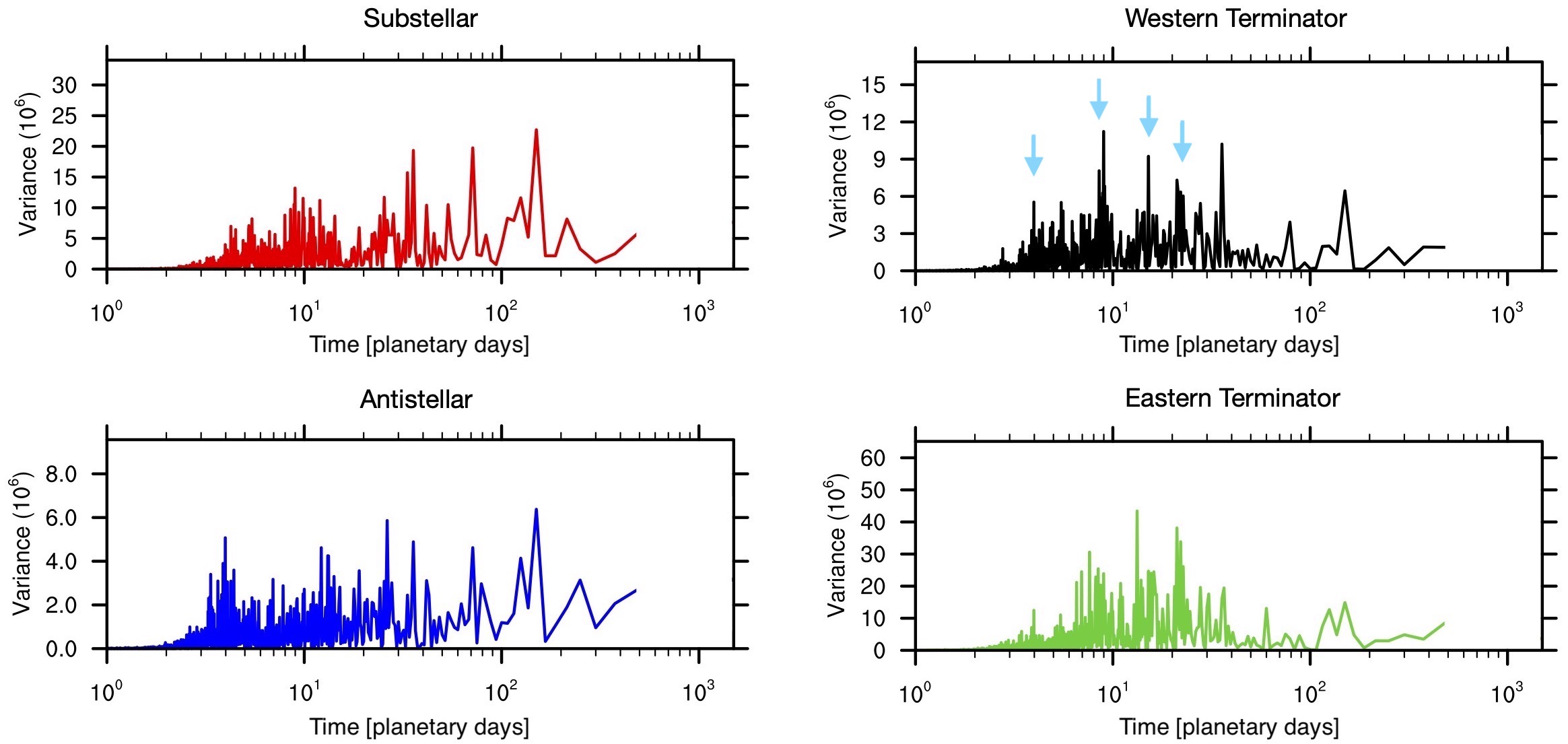}}
  \caption{Power spectrum normalised by the variance of the disc-averaged
    temperature flux time-series in Fig.~\ref{fig:flux_long}.  All four spectra
    are broad and ``rich'' (i.e. many peaks over a wide range of periods).  
    All of the spectra generally contain peaks at $\sim$8-planetary days and 
    $\sim$15-planetary days, with peaks at
    shorter period ($\sim$4-planetary days) and longer period
    ($\sim$25-planetary days): these are labelled in the upper right frame.
      The corresponding four spectra at
    high $p$-levels in the deep atmosphere simulations (not shown) are markedly
    different -- as might be expected from Fig.~\ref{fig:deep_flux}; see also
    \citet{Choetal21}, Fig.~4.  Spectra in the deep region are not broad and
    contain only few peaks.  They are also essentially identical to each other,
    showing that the life-cycles in the deep region are much more regular and
    differ only in phase.  Many peaks can be directly related to the temperature
    redistribution by the modons, as in Fig.~\ref{fig:flux_long}.}
  \label{fig:timeseries_spectra}
\end{figure*}

The sequence of frames shown in Fig.~\ref{fig:temp_flux}a follows the modon from
its formation near the sub-stellar point ($t = 93$), to its cooling ($t =
\{98,100\}$), to breakup and diffusion ($t = 102$) and the subsequent
reformation and re-heating ($t = 105$).  Overall, the modon’s periodic motion in
this example redistributes planetary-scale regions of hot {\it and} cold air
zonally (both westward) with $\sim$12-day period.  The cyclonic modon initially
sequesters heat (hot temperature) at the day-side in each of its cyclones and
form {\it two} distinct hot patches straddling the equator (separated by a
narrow, meandering, cold jet between them).  As already discussed, meridional
redistribution occurs as well ($t = \{100,102\}$).  Note that, the cyclonic
modon transports cold air from the night-side ($t = 102$); at the same time, the
anti-cyclonic modon can also be seen transporting hot air.  Shortly after, the
once cold cyclonic modon has warmed up upon reaching the sub-stellar point ($t =
105$).

\begin{figure*}%
  \vspace*{1cm}  
  \centerline{\includegraphics[scale=.15]{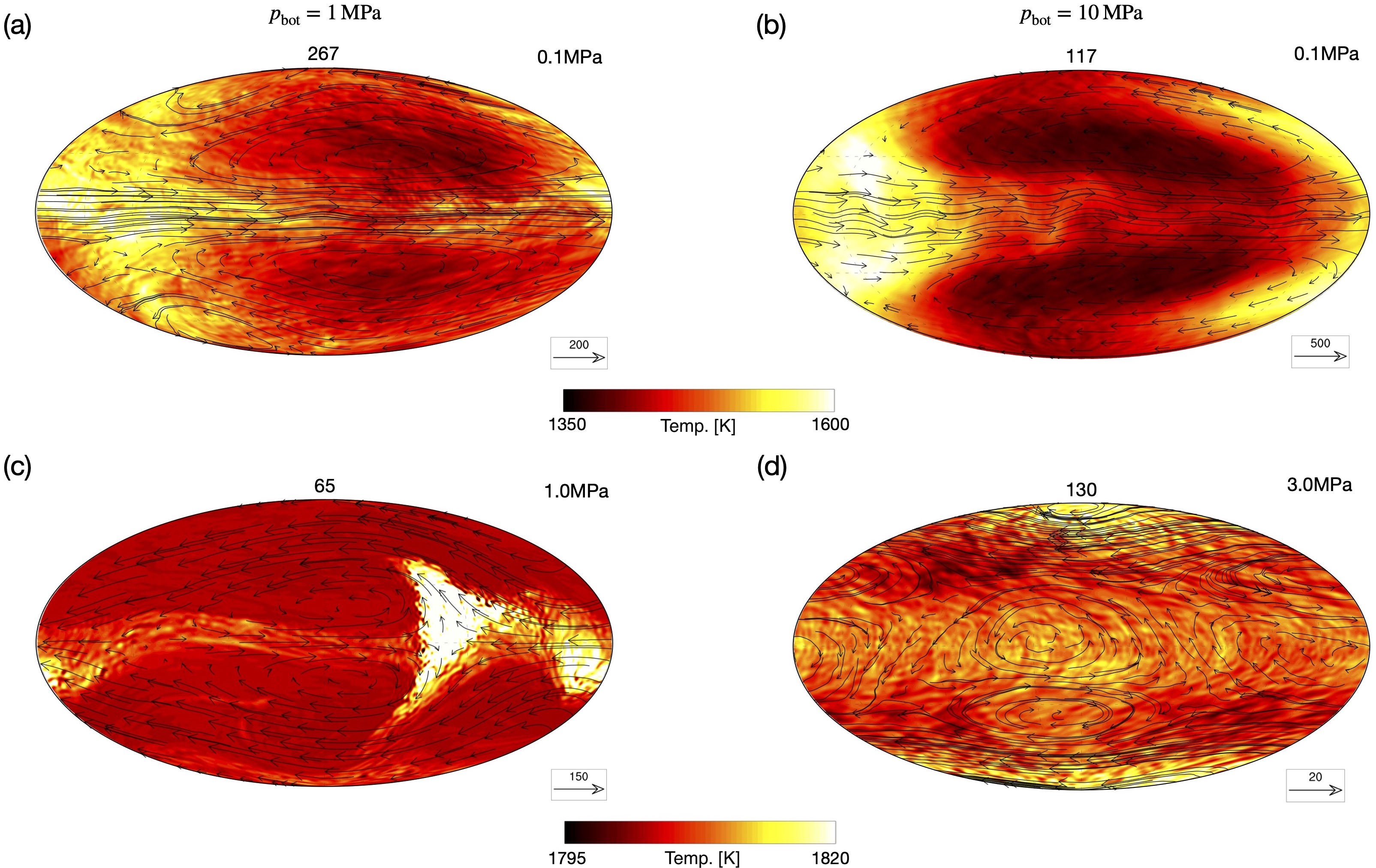}}
  \caption{Temperature and wind (black arrows) fields centred on the night-side,
    from two T170L200 deep atmosphere simulations with $p_{\rm bot} \in \{1,
    10\}$; $\mathfrak{p} = 8$, $\Delta t = 8.0\!\times\!  10^{-5}$ and $\nu_{16}
    = 10^{-38}$.  Time is indicated above each frame.  The reference vector
    length and temperature range are adjusted to accommodate the different
    background (flow and temperature) conditions.  The four frames illustrate
    modons occurring at different times and $p$-levels (as labelled) throughout
    both simulations.  As in the shallow atmosphere simulations
    (cf. Fig.~\ref{fig:temp_flux}), modons mix and redistribute hot (and cold)
    regions over the planet -- including in region that is not subject to direct
    thermal forcing.  All of the modons at the centre of each frame are cyclonic
    modons.}
  \label{fig:deep_temp}
\end{figure*}

The modon’s rearrangement of temperature leads to discernible signatures in the
time-series of the disc-integrated flux at each of the four chosen locations, as
shown in Fig.~\ref{fig:temp_flux}b.  In general, the SS flux (at the peak of the
secondary eclipse) is the largest and the AS flux (at the peak of the primary
eclipse) is the smallest of the four fluxes.  But, the variations in each flux
is such that SS flux can be lower than the ET flux and the ET flux can be lower
than the WT flux, at different times and with different periods.
Fig.~\ref{fig:temp_flux}a and Fig.~\ref{fig:temp_flux}b both show that the
hottest (coldest) region can be located over a broad range of longitude, and on
either side of the sub-stellar (anti-stellar) point.

Fig.~\ref{fig:flux_long} shows the flux time-series at $p = 0.1$ over a long
duration from a simulation at T341L20 resolution with $\mathfrak{p} = 8$,
$\nu_{16} = 1.5\!\times\! 10^{-43}$ and $\Delta t = 4\!\times\! 10^{-5}$; this
is the simulation presented in Fig.~\ref{fig:breakup}.  Top and bottom frames in
Fig.~\ref{fig:flux_long} show the fluxes at the sub-stellar and anti-stellar
points (top) and the fluxes at the two terminators (bottom), respectively.  The
generic flux variability seen in Fig.~\ref{fig:temp_flux}b, associated with the
modon's generally cyclic behaviour, persists for $1500$~planetary days; hence,
the simulations and behaviours presented in this paper are assumed to be in
quasi-equilibration.  We note that this simulation is also numerically converged
\citep[see][]{SkinCho21}.  Although not quantitatively comparable, {\it
  transitions} to several different quasi-equilibrium states are expected, based
on extremely long-duration (over 80,000 planetary days)
simulations at much lower resolution \citep{Thra11} -- hence, the ``quasi''
qualification (because of the transitions).  Roughly, two states are readily
apparent in the figure: ``active'' state and ``quiet'' state, based on period
and/or amplitude. Note that these temperature time-series track the
flow's global eddy kinetic energy (cf. Fig.~\ref{fig:EKE} in the Appendix).
The quiet regions in the fluxes correspond to the ``omega
blocking'' quadrupolar vortex configuration which rapidly mixes the hot and cold
regions more evenly though out the flow.  The bursts in the flux are caused by
the slower moving modons, which do not mix the atmosphere as quickly and carry
the heat (hot temperature) as they migrate.  The slower moving modon traps air
on the day-side for longer period, causing the flux amplitude to be higher on
the day side.  This is also why the SS and ET flux variances are high, compared
to the other fluxes, during the modon formation phase (see also
Fig.~\ref{fig:temp_flux}b).

Fig.~\ref{fig:timeseries_spectra} shows the power spectra of the disc-averaged
flux time-series in Fig.~\ref{fig:flux_long}.  Information such as this (from
converged simulations) can aid in planning or scheduling observations
\citep{Choetal21,SkinCho21}.  The spectra are computed from the time-series data
between days $25$ and $1500$ to neglect variability associated with the
simulation's initial ramp-up phase.  The power spectra are normalised such that
the variance of the de-trended series is equal to $\sum_j S(j)\,\d f$, where
$S(j)$ is the time-series data with $j$ samples and $\d f$ is the frequency
spacing (4~per day in this case).

All of the power spectra in Fig.~\ref{fig:timeseries_spectra} are broad and
``rich'', with many peaks over a wide range of periods.  In general, all the
spectra contain peaks at $\sim\! 8$-planetary days and $\sim\!
15$-planetary days.  The spectra also contain peaks of shorter
period ($\sim\! 4$-planetary days) and longer period ($\sim\!
25$-planetary days), including those of very long periods
(i.e. $\gtrsim 100$-planetary days).  All of these periods can be
directly attributed to flow induced temperature field evolutions discussed in
this work.  Further, as can be surmised from data shown below, the corresponding
four spectra at high $p$-levels in the deep atmosphere simulations are markedly
different than those in Fig.~\ref{fig:timeseries_spectra}: the spectra from the
high $p$-levels are not broad and contain only few peaks
\citep[see][]{Choetal21}.  These levels, of course, do not exist in shallow
atmosphere simulations.  The corresponding four spectra at high $p$-levels are
also essentially identical to each other, showing that the life-cycles in the
deep regions are much more regular and differ only in phase, since phase
differences are not detected by a power spectrum \citep{Choetal21}.

Fig.~\ref{fig:deep_temp} presents the temperature $T$ and wind fields of two
deep atmosphere simulations which include $p$-levels that are not thermally
forced (i.e. $ p \ge 1$).  These simulations are identical in setup, with
T170L200 resolution, $\mathfrak{p} = 8$, $\nu_{16} = 10^{-38}$ and $\Delta t =
8\!\times\!  10^{-5}$ -- except for the different values of $p_{\rm bot} = \{1,
10\}$.  Note that the two simulations presented are not numerically converged
\citep[see][]{SkinCho21}; however, they are shown here are for illustrative
purposes -- in order to demonstrate that modons form in simulations with
different $p_{\rm bot}$ values, at a variety of $p$-levels and simulation times.
These modons are still expected to form in converged (i.e. $\ga$ T341L4000) deep
atmosphere simulations.  Moreover, the simulations presented are still greater
in resolution than nearly all past extrasolar planet studies to date.  As in
Fig.~\ref{fig:temp_flux}a, the $T$ field is displayed in the Mollweide
projection, but it is now centred on the anti-stellar point to show modons
located on the planet's night-side.  Note also, the reference vector lengths in
the different frames are adjusted to accommodate the different background (flow
and temperature) conditions at the $p$ levels presented.

As can be seen in Fig.~\ref{fig:deep_temp}, modons occur throughout the modelled
atmospheres -- independent of the location of the domain bottom $p_{\rm bot}$.
As just alluded to, for a {\it quantitatively} accurate assessment
(e.g. detailed comparison with converged deep simulations), a sufficient number
of vertical levels spanning the domain range is required.  However, the modons
in Fig.~\ref{fig:deep_temp} also redistribute hot and cold patches of air as
they translate around the planet, periodically breaking and reforming into
several different configurations of vortices -- as in the shallow atmosphere
simulations (cf. Fig~\ref{fig:temp_flux}).  Frames~(a) and~(b) shows explicitly
that cyclonic modons also sequester and transport cold air.  Note that the
equatorial jet, which separate the cyclones of the modon, is warmer than the
imposed $T_{\rm eq}$ in both frames.  Frames~(c) and~(d) show that modons, which
form in $p$-regions not directly forced, transport and redistribute temperature.
Not surprisingly, in these frames modons induce hottest patches of air to be at
the night-side -- and even at the poles (Fig.~\ref{fig:deep_temp}d).  Such
features are significant because, although much lower in amplitude than in the
lower $p$-levels, fluxes can still come from the higher $p$-levels shown.

\begin{figure*}%
  \vspace*{0.2cm}
  \centerline{\includegraphics[scale=.175]{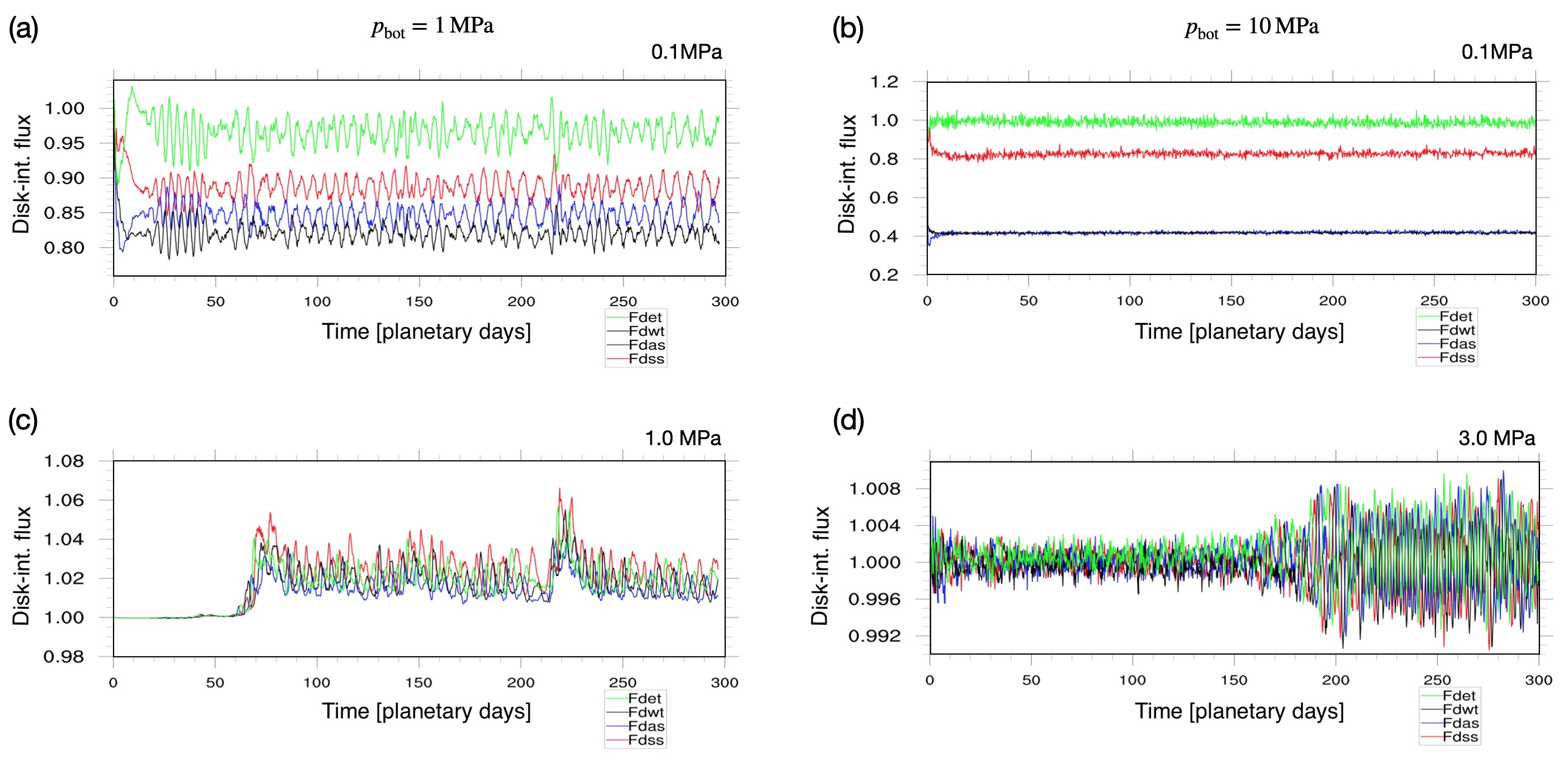}}
  \caption{Disc-integrated temperature flux for the simulations presented in
    Fig.~\ref{fig:deep_modons}.  Similarly to the shallow atmosphere
    simulations, the modon's effect on the global temperature distribution
    produces periodic signatures in the time-series of the disc-integrated
    fluxes at each of the planet's key locations and for different values of
    $p_{\rm bot}$ (here $\{1,10\}$).  In general, the flux variations are
    largest near the top of the atmosphere however periodic signals associated
    with the modons movements also occur at deeper levels.  Quantitatively, the
    time-series behaviours are different compared to that in the shallow
    atmosphere (Fig.~\ref{fig:flux_long}), consistent with the difference in the
    flows (cf. Figs.~\ref{fig:modon} and \ref{fig:deep_temp}); however, features
    such as variability, quasi-equilibration and multiple states within and
    across $p$-levels are characteristic of both shallow and deep atmospheres.
    Strong vertical coupling is also evident in (a) and (c) -- e.g. at $t
    \approx$ 60, 165 and 215 (see text).}
  \label{fig:deep_flux}
\end{figure*}

Fig.~\ref{fig:deep_flux} shows the corresponding flux time-series at several key
locations on the planet for the simulations presented in
Fig.~\ref{fig:deep_temp} in the time interval $t = [0, 300]$.  As in
Fig.~\ref{fig:deep_temp}, left and right columns correspond to simulations where
the bottom of the domain $p_{\rm bot}$ is placed at 1 and 10, respectively.  Top
and bottom rows show the fluxes in the radiatively forced region
(Fig.~\ref{fig:deep_flux}a and Fig.~\ref{fig:deep_flux}b) and below it
(Fig.~\ref{fig:deep_flux}c and Fig.~\ref{fig:deep_flux}d), respectively.  The
figure shows that the generic behaviour seen in the shallow atmosphere
simulations (e.g. variabilities at multiple time-scales, state transitions and
reduction in flux amplitude with $p$), also occurs in the deep atmospheres.
Again, these behaviours are directly attributable to the behaviour of modons.
The dynamism, however, requires that both the vertical and horizontal
resolutions are sufficient to capture the modons with reasonable accuracy.

More quantitatively, the signals are the strongest at the eastern terminator
location at $p = 0.1$.  This is in contrast with the fluxes from the shallow
atmosphere simulation (cf. Fig.~\ref{fig:flux_long}), which generally exhibits
the highest amplitude signals at the sub-stellar location.  Hence the signal is
eastward shifted in the bulk (Fig.~\ref{fig:deep_flux}a); and, the shift is more
pronounced with higher $p_{\rm bot}$ value (Fig.~\ref{fig:deep_flux}b).  This is
due to the lower horizontal resolution, as well as inadequate vertical
resolution for the $p_{\rm bot}$ values (Skinner and Cho, in prep.).  As
discussed above, a robust $p_{\rm bot} = 10$ simulation requires $\sim$\,4000
$p$-levels at T341 resolution.  Qualitatively, a common feature in the deep
atmosphere fluxes is a sudden jump in amplitude (e.g. at $t\sim 70$ in
Fig.~\ref{fig:deep_flux}c and $t\sim 190$ in Fig.~\ref{fig:deep_flux}d).  And,
the jumps are generally correlated with sudden changes at $p = 0.1$
\citep{Choetal21}.  Such behaviour suggests the importance of properly resolving
the directly forced $p$-region in deep atmosphere simulations and the utility of
accurately modelling shallow atmospheres as well, since high horizontal
resolution is still necessary even if high vertical resolution is achieved.

\section{Conclusions}\label{conclusion}

In this paper, we have summarised the results from a large set of
high-resolution (up to T682L20 and T341L200), focusing on numerically converged
simulations aimed at carefully studying the dynamics in a tidally synchronised
extrasolar planet atmosphere.  We have found that a generic solution in the
implemented setup is a pair of ``oppositely-signed'' (cyclonic and
anti-cyclonic) planetary-scale modons. Note that, in general, the
cyclonic modon is stronger than the anti-cyclonic modon; hence, the former tends
to exert a stronger influence on the overall flow (and temperature
redistribution) -- although both modons are important.  Both modons are also
strongly barotropic and form at a variety of depths down to the $p = 10$\,MPa
level with different vertical extents, depending on the location of the domain
bottom level $p_{\rm bot}$ \citep{SkinCho21}.

The modons in synchronised atmospheres are highly dynamic and feature a complex
set of non-linear motions.  Such behaviours have not been previously reported.
Crucially, the complex motions are a result of non-linear interactions with
energetic small-scale eddies and waves, which are not accurately captured in
low-resolution and/or low-order dissipation simulations.  For example, in our
high-resolution simulations with high-order dissipation, we have found that the
modons are surrounded by a very large number of small-scale vortices and waves
that develop in response to the modons' attempt to adjust in the ageostrophic
flow environment.  Ageostrophy is a generic characteristic of hot synchronised
planet atmospheres.  Thus, these atmospheres are highly turbulent, which is
laterally extremely anisotropic (i.e. very ``patchy'' level-wise).  The
quasi-periodic nature of the flows also means the turbulence is robustly
maintained overall, as the atmosphere is repeatedly stirred.

Broadly, our results present several significant implications for current and
future observations of extrasolar planet atmospheres.  In particular, we find
that the atmospheres exhibit multiple equilibrium states, associated with the
different cyclic behaviour of modons.  The states produce temperature flux
signatures at different locations over the planet.  In different life-cycles,
the modons undergo changes in direction, size and strength.  The cyclonic and
anticyclonic modons generally behave differently.  In particular, the cyclonic
modon typically dissipates completely near the eastern terminator before a new
modon forms and the cycle repeats, whereas the anti-cyclonic modon often
separates and the constituent anti-cyclones move towards their
respective poles.  Such motions result in vigorously mixing of temperature (and
active species) in the atmosphere {\it on the planetary-scale} -- by storms,
rather than by a zonal equatorial jet.

\section*{Acknowledgements}

The authors thank Michael E. McIntyre, Heidar Th. Thrastarson and Inna
Polichtchouk for helpful discussions.  We are grateful for the hospitality of
James Stone and the Department of Astrophysical Sciences, Princeton University,
where some of this work was completed.  J.W.S. is supported by UK's Science and
Technology Facilities Council research studentship.  We thank the referee
for the comments.

\section*{Data Availability}
The data underlying this article will be shared on reasonable request to the
corresponding author.

\appendix 
\section*{Appendix}
\renewcommand{\thefigure}{A\arabic{figure}}
\setcounter{figure}{0}

\begin{figure}
  \vspace*{2cm}
  \centerline{\includegraphics[scale=.15]{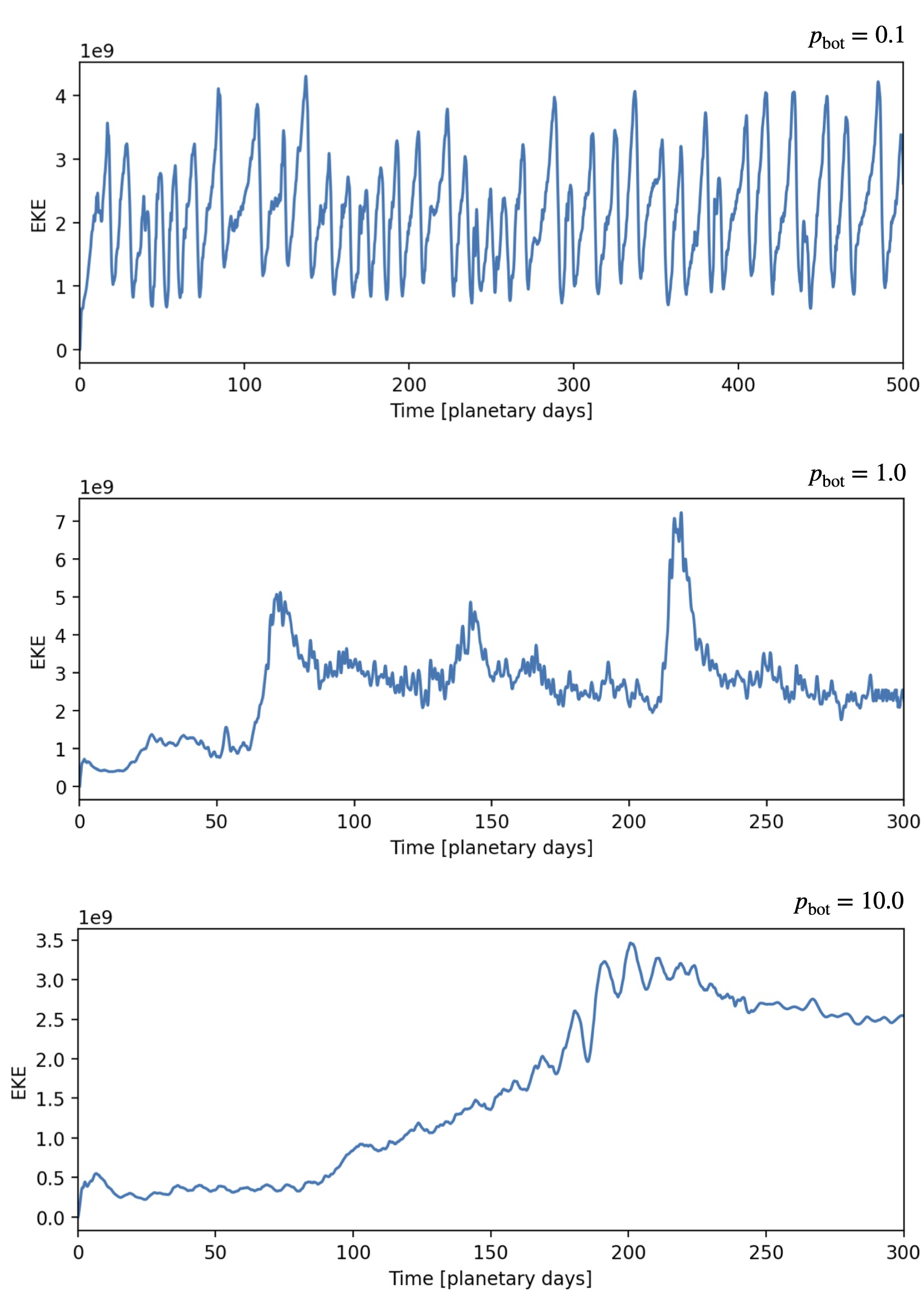}}
  \caption{Global eddy kinetic energy time-series for the simulations
    presented in this paper.  The shallow atmosphere (i.e. $p_{\rm bot} = 0.1$)
    simulations equilibrate after $\sim$10 days. The deep atmosphere simulations
     (i.e. $p_{\rm bot} = \{1.0, 10.0\}$) show a ``jump behaviour'' consistent with 
     the temperature time-series (c.f. Fig~\ref{fig:deep_temp}) suggesting multiple
     ``quasi-equilibrium'' states over long \citep{Thra11} as well as short 
     \citep{Choetal21} time-scales. }
  \label{fig:EKE}
\end{figure}

\begin{thebibliography}{9}

\bibitem[\protect\citeauthoryear{Abramowitz \& Stegun}{1965}]{AbraSteg65}
  Abramowitz M., Stegun I. A., 1965, Handbook of Mathematical Functions: with
  Formulas, Graphs, and Mathematical Tables, revised ed., Dover, New York

\bibitem[\protect\citeauthoryear{Anstey et al.}{2013}]{Anstetal13} Anstey J. et
  al., 2013, J. Geophys. Res. Atmos., 118, 3956

\bibitem[\protect\citeauthoryear{Armstrong et al.}{2017}]{Armstrong17} 
Armstrong D.J., De Mooij E., Barstow J., Osborn H.P., Blake J., Fereshteh Saniee N.,
2017, Nature, 1, 0004.

\bibitem[\protect\citeauthoryear{Asselin}{1972}]{Asse72} Asselin R.,
  1972, Mon. Wea. Rev., 100, 487

\bibitem[\protect\citeauthoryear{Boyd}{2000}]{Boyd00} Boyd J. P., 2000,
  Chebyshev and Fourier Spectral Methods, 2nd ed., Dover, New York

\bibitem[\protect\citeauthoryear{Byron \& Fuller}{1992}]{ByroFull92} Byron,
  F. W., Fuller, R. W., 1992, Mathematics of Classical and Quantum Physics,
  Dover, New York

\bibitem[\protect\citeauthoryear{Canuto et al.}{1988}]{Canuto88} Canuto C.,
   Hussaini M. Y., Qarteroni A., Zang T. A., 1988, Spectral Methods in Fluid
   Dynamics (New York, NY: Springer)

\bibitem[\protect\citeauthoryear{Cho and Polvani}{1996a}]{ChoPol96a} Cho J. Y-K.,
   Polvani L., 1996, Phys. Fluids, 8, 1531

\bibitem[\protect\citeauthoryear{Cho and Polvani}{1996b}]{ChoPol96b} Cho J. Y-K.,
   Polvani L., 1996, Science, 273, 5273   

\bibitem[\protect\citeauthoryear{Cho et al.}{2003}]{Choetal03} Cho
  J. Y-K., Menou K., Hansen B. M. S., Seager S., 2003, ApJ, 587, L117

\bibitem[Cho(2008)]{Cho08} Cho, J. Y-K. 2008, Phil.\ Trans.\ R.\ Soc.\
  A, 366, 4477

\bibitem[\protect\citeauthoryear{Cho et al.}{2008}]{Choetal08} Cho
  J. Y-K., Menou K., Hansen B. M. S., Seager S., 2008, ApJ, 675, 817

\bibitem[\protect\citeauthoryear{Cho, Polichtchouk \&
    Thrastarson}{2015}]{Choetal15} Cho J. Y-K., Polichtchouk I., Thrastarson
  H. Th., 2015, MNRAS, 454, 3423

\bibitem[\protect\citeauthoryear{Cho et al.}{2019}]{Choetal19}  Cho J. Y-K. et
  al., 2019, Exoplanets and the Sun in Galprin, B. \& Read, P. L. eds., Zonal
  Jets: Phenomenology, Genesis, and Physics. Cambridge University Press,
  Cambridge, p. 550

\bibitem[\protect\citeauthoryear{Cho, Skinner \& Thrastarson}{2021}]{Choetal21}
  Cho J. Y-K., Skinner J. W., Thrastarson H. Th., 2021, ApJL, 913, L32 

\bibitem[\protect\citeauthoryear{Cooper \& Showman}{2005}]{Cooper05}
  Cooper C. S., Showman A. P., 2005, Dynamics and disequilibrium
  carbon chemistry in hot Jupiter atmospheres, with application to
  HD~209458b, ApJ, 629, L45--L48

\bibitem[\protect\citeauthoryear{Dang et al.}{2018}]{Dang18} 
Dang, L., et al., 2018., Nature, 2, 220–227. 

\bibitem[\protect\citeauthoryear{Durran}{2010}]{Durr10} Durran D. R., 2010,
  Numerical Methods for Fluid Dynamics with Applications to Geophysics, 2nd ed.,
  Springer, New York

\bibitem[\protect\citeauthoryear{Eliasen et al.}{1970}]{Eliaetal70} Eliasen E.,
  Mechenhauer B., Rasmussen E., 1970, Copenhagen Univ., Inst. Teoretisk
  Meteorologi, Tech.  Rep. 2

\bibitem[\protect\citeauthoryear{Gill}{1980}]{Gill80} Gill A.E., 1980,
  Q. J. Roy. Met. Soc., 106, 447

\bibitem[\protect\citeauthoryear{Grillmair et al.}{2008}]{Grill08} 
Grillmair, C. J., et al., 2008., Nature, 456, 767–769.

\bibitem[\protect\citeauthoryear{Holton}{2004}]{Holt04} Holton J. R., 2004,
  An Introduction to Dynamic Meteorology, 4th ed., Academic Press, San Diego

\bibitem[\protect\citeauthoryear{Heng et al.}{2011}] {Hengetal11} Heng K.,
  Frierson D., Phillipps P., 2011, MNRAS, 418, 4, 2669–2696

\bibitem[\protect\citeauthoryear{Hogg \& Stommel}{1985}]{hoggstom85} Hogg N. G.,
  Stommel H. M., 1985, Proc. R. Soc. Lond. A, 397, 1


\bibitem[\protect\citeauthoryear{Jackson et al.}{2019}]{Jackson19}
  Jackson B., Adams E., Sandidge W., Kreyche S., Briggs J., 2019, ApJ, 157, 239

\bibitem[\protect\citeauthoryear{Jeffreys}{1925}]{Jeffreys1925} Jeffreys H.,
  1925, Proc. R. Soc. London, Ser. A, 107(742), 189–206.

\bibitem[\protect\citeauthoryear{Kizner}{2006}]{kizner06} Kizner Z., 2006,
  Phys. Fluids, 18, 5

\bibitem[\protect\citeauthoryear{Lahaye \& Zeitlin}{2012}]{LahaZeit12}
  Lahaye N., Zeitlin V., 2012,  J. Fluid Mech. 706, 71

\bibitem[\protect\citeauthoryear{Liu \& Showman}{2013}]{LiuShow13} Liu
  B., Showman A. P., 2013, ApJ, 770, 42

\bibitem[\protect\citeauthoryear{Matsuno}{1966}] {Mats66} Matsuno T., 1966,
  J. Meteorol. Soc. Japan, 44, 25
  
\bibitem[Menou \& Rauscher(2009)]{Menou09} Menou K., 
  Rauscher E., 2009, ApJ, 700, 887

  \bibitem[Mendon\c{c}a (2020)]{Mend20} Mendon\c{c}a J. M., 2020, MNRAS, 491, 1

\bibitem[\protect\citeauthoryear{Orszag}{1970}]{Orsz70} Orszag A.,
  1970, J. Atmos. Sci., 27, 890


\bibitem[\protect\citeauthoryear{Pedlosky}{1987}]{Pedl87} Pedlosky J. R., 1987,
  Geophysical Fluid Dynamics, 2nd ed., Springer-Verlag, New York

  \bibitem[\protect\citeauthoryear{Polichtchouk and Cho}{2012}] {PoliCho12}
    Polichtchouk I., Cho J. Y-K., 2012, MNRAS, 424, 2, 1307–1326

\bibitem[\protect\citeauthoryear{Polichtchouk et al.}{2014}] {Polietal14}
  Polichtchouk I., Cho Y-K. J., Watkins C., Thrastarson H. Th., Umurhan O. M.,
  Juarez M. T., 2014, Icarus, 229, 355

\bibitem[\protect\citeauthoryear{Rauscher \& Menou}{2010}]{Rauscher10} Rauscher
  E., Menou K., 2010, ApJ, 714, 1334

\bibitem[\protect\citeauthoryear{Rex}{1950}] {Rex50} Rex D. F., 1950, Tellus,
  2:4, 275

\bibitem[\protect\citeauthoryear{Rivier, Loft \& Polvani}{2002}] {Rivietal02}
  Rivier L., Loft R., Polvani L. M., 2002, Mon. Wea. Rev., 130, 1384

\bibitem[\protect\citeauthoryear{Robert}{1966}]{Robe66} Robert A.,
  1966, J. Met. Soc. Japan, 44, 237

\bibitem[Salby (1996)]{Salb96} Salby M. L., 1996, Fundamentals of Atmospheric
  Physics, Academic Press, San Diego

\bibitem[\protect\citeauthoryear{Scott et al.}{2004}]{Scotetal04}
  Scott R. K., Rivier L., Loft R., Polvani L. M, 2004, NCAR Technical
  Note No.\ 456 

\bibitem[Showman et al.(2008)]{Showmanetal08a} Showman A. P., 
  Cooper C. S., Fortney J. J., Marley M. S., 2008, ApJ, 682, 559

\bibitem[Showman et al.(2008)]{Showmanetal08b} Showman A. P., 
  Menou K., Cho, J. Y-K., 2008, in ASP Conf. Ser. 398, Extreme 
  Solar Systems, ed. D. Fischer et al. (San Francisco, CA: ASP)

\bibitem[\protect\citeauthoryear{Showman \& Guillot}{2002}]{Showman02} Showman
  A. P., Guillot T. 2002, A\&A, 385, 166-180

\bibitem[\protect\citeauthoryear{Showman \& Polvani}{2011}]{ShowPol11}
  Showman A. P., Polvani L. M., 2011, ApJ, 738, 71 

\bibitem[\protect\citeauthoryear{Skinner \& Cho}{2021}]{SkinCho21} Skinner
  J. W., Cho J. Y-K., MNRAS (in press); arXiv:2010.09695

\bibitem[\protect\citeauthoryear{Stern}{1975}]{Ster75}
   Stern M. E., 1975, J. Mar. Res. 33, 1

\bibitem[Strikwerda (2004)]{Stri04} Strikwerda J. C., 2004, Finite Difference
  Schemes and Partial Differential Equations, 2nd ed., Society for Industrial
  and Applied Mathematics, Philadelphia

\bibitem[Thrastarson \& Cho(2010)]{ThraCho10} Thrastarson H.Th., Cho,
  J.Y-K. 2010, Apj, 716, 144

\bibitem[\protect\citeauthoryear{Thrastarson}{2011}]{Thra11}
  Thrastarson H. Th., 2011, PhD. Thesis, Queen Mary University of London

\bibitem[\protect\citeauthoryear{Thrastarson \& Cho}{2011}]{ThraCho11}
  Thrastarson H. Th., Cho J. Y-K., 2011, ApJ, 729, 117

\bibitem[\protect\citeauthoryear{Vallis}{2017}]{Vall17} Vallis G. K., 2017,
  Atmospheric and Oceanic Fluid Dynamics, 2nd ed., Cambridge University Press,
  Cambridge.

\bibitem[\protect\citeauthoryear{Woolings et al.}{2018}]{Wooletal18}
  Woolings T., et al., 2018, Cur. Cli. Change Rep., 4, 287
  
\bibitem[\protect\citeauthoryear{Wu, Sarachik \& Battisti}{2001}]{Wuetal01}
  Wu Z., Sarachik E. S., Battisti D.S., 2001, J. Atmos. Sci., 58, 724

\bibitem[\protect\citeauthoryear{Zellem et al.}{2014}]{Zellem14}
Zellem, R. T., et al., 2014. ApJ., 790, 53–62.

\bibitem[\protect\citeauthoryear{Zhang et al.}{2018}]{Zhang18}
  Zhang, M., et al., 2018, A.J , 155, 83 

\end{thebibliography}
\end{document}